\documentclass[12pt]{elsarticle}
\usepackage{hyperref}
\usepackage{graphicx}

\journal{Applied Surface Science}

\begin{document}
\begin{frontmatter}
\title{Laser-induced ablation of metal in liquid}

\author[1,3]{Yu.V.~Petrov}\ead{uvp49@mail.ru}
\author[1]{V.A.~Khokhlov}
\author[2,1]{V.V.~Zhakhovsky}
\author[1,2]{N.A.~Inogamov}
\address[1]{Landau Institute for Theoretical Physics of RAS, 1-A Akademika Semenova av., Chernogolovka, Moscow Region, 142432, Russia}
\address[2]{Dukhov Research Institute of Automatics, 22 Sushchevskaya st., Moscow, 127055, Russia}
\address[3]{Moscow Institute of Physics and Technology, 9 Institutskii per., Dolgoprudny, Moscow Region, 141700, Russia}



\begin{abstract}
 Laser ablation in liquid (LAL) is important perspective way to compose nanoparticles (NP) necessary for modern technologies.
 LAL is not fully understood.
 Deep understanding is necessary to optimize processes and decrease high price of the LAL NPs.
 Today there are two groups of studies: in one of them scientists go from analyzing of bubble dynamics (thus they proceed from the late stages),
   while in another one scientists investigate early stages of ablation.
 In the present paper we consider the process as whole: from ablation and up to formation of a bubble and its inflation.
 Thus we cover extremely wide range of spatiotemporal scales.
 We consider role of absorbed energy and duration of pulse (femtosecond, multi-picosecond, nanosecond).
 Importance of supercritical states is emphasized.
 Diffusive atomic and hydrodynamic mixing due to Rayleigh-Taylor instability and their mutual interdependence are described.

 Liquid near contact with metal is heated by dissipation in strong shock and due to small but finite heat conduction in liquid;
   metal absorbing laser energy is hot and thus it serves as a heater for liquid.
 Spatial expansion and cooling of atomically mixed liquid and metal causes condensation of metal into NPs when pressure drops below critical pressure for metal.
 Development of bubble takes place during the next stages of decrease of pressure below critical parameters for liquid and below ambient pressure in liquid.
 Thin hot layer of liquid near contact expands in volume to many orders of magnitude filling the inflating bubble.
\end{abstract}

\end{frontmatter}

%
\noindent{\it Keywords}: laser ablation in liquid, overcritical states, water adiabatic curves
%
%
%
%

\begin{figure}       
   \centering   \includegraphics[width=0.75\columnwidth]{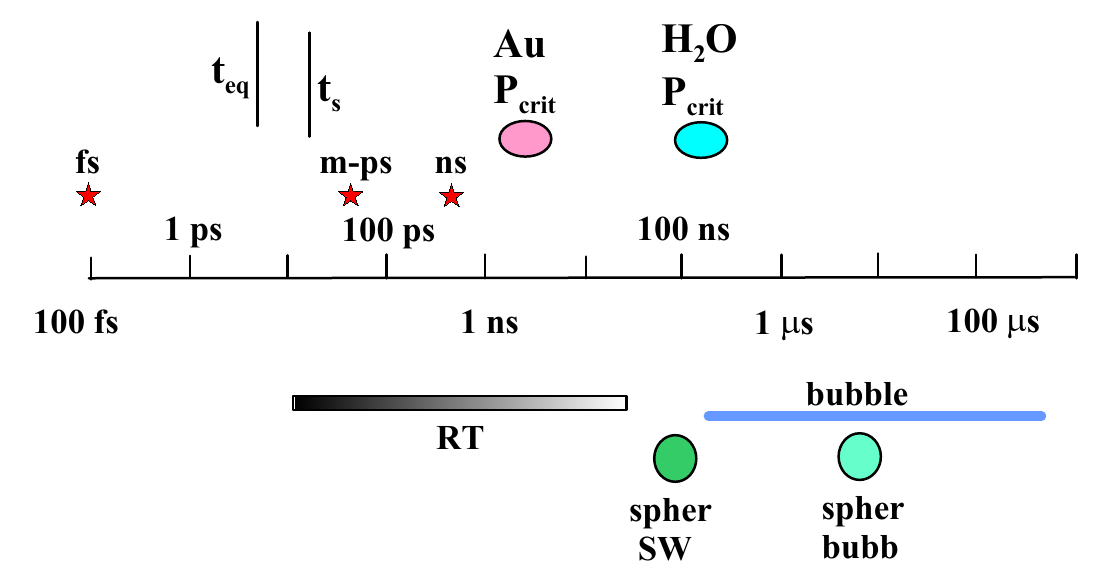}
\caption{\label{ris:01-hierarchy}  The hierarchy of processes triggered by laser action.
The stars mark durations $\tau_L=$ 100 fs (fs), 50 ps (m-ps, multi-picosecond), and 0.5 ns (ns) considered in paper.
Range of absorbed fluences is $F_{abs}=0.3-1$ J/cm$\!^2,$ radius of a laser beam at surface of metal target immersed in liquid is $R_L\sim 100$ $\mu$m.
The duration $t_{eq}$ for gold separates values of $\tau_L$ when we have to include two-temperature effects $T_e\gg T_i,$ see \cite{INA.jetp:2018.LAL}.
During the stage lasting time $t_{eq}$ equilibration of electron and ion temperatures $T_e$ and $T_i$ takes place \cite{INA.jetp:2018.LAL}.
If $\tau_L < t_s=d_T/c_s$ (stress-confinement) then the largest ratios $p/F_{abs}\sim 1/d_T$ are achieved; here $d_T$ is thickness of a heat affected zone,
  $c_s$ is speed of sound, $p$ is the maximum pressure created at the end of absorption of a laser pulse.
The ellipses mark stages when contact pressure decreases below critical pressure of gold and liquid (in our paper this is water).
Bubble begins to form and exists during the range of times marked as "bubble."
The range "RT" refers to Rayleigh-Taylor instability (RTI). This is a hydrodynamic instability mixing gold and water.
Here the beginning of the RT range is shown for the case of femtosecond action.
For longer pulses a role of RTI in mixing and production of nanoparticles decreases.
    }  \end{figure}

\section{Introduction}


 There are two competing approaches to production of nanoparticles (NP, NPs).
 They are chemical and laser approaches \cite{Stephan:2017.review,XIAO:2017}.
 NPs are used in many important scientific and industrial applications \cite{Stephan:2017.review,XIAO:2017}.
 Chemical production is cheap but laser fabrication is more simple, clean, and more green in an ecological sense \cite{Stephan:2017.review,XIAO:2017}.
 Therefore studies directed to understand and on this base to optimize and to do cheaper production of NPs by laser ablation in liquid (LAL) are valuable.
 But today in spite of significant efforts many aspects of LAL remains unclear.
 This is because it is difficult to describe theoretically different and closely related physical processes proceeding at a wide logarithmical temporal range, see Fig. \ref{ris:01-hierarchy}.
 Also early stages, before formation of a bubble, remains experimentally unexplored; but see paper \cite{Bulgakov:2017}
    where attempt has been made to follow state of an irradiated surface during laser action.
 In \cite{Bulgakov:2017} (in their inset in Fig. 9(a)) authors see decrease of reflection from the gold surface in water during a nanosecond laser pulse.


 Short lasting laser action strongly changes thermal and mechanical state of a system consisting from an absorbing target and transparent liquid or solid media surrounding a target;
    below we consider gold target contacting with water or glass.
 Absorption of light and heating of gold through transparent media is concentrated in an extremely thin surface disk; therefore a small amount of NPs is produced by a single shot.
 Radius of the disk is defined by radius of a laser beam $R_L$ (usually $R_L\sim 0.1$ mm), while its thickness equals to thickness of a heat affected zone $d_T$ and is a fraction of micron.
 Further, after finishing absorption, a long multi-step process of relaxation follows.
 Positive and negative momentum is carried away from the surface layer by non-linear acoustic waves (shocks in water and in gold).
 While the heat affected zone evolves separately (from the shocks) with velocities which are small relative to speed of sound.


 Today detailed data concerning late stages beginning from $\sim 1$ $\mu$s are collected \cite{Stephan:2017.review,XIAO:2017,Amans:2016:APL}.
 These data includes experimental measurements of a trajectory $R(t)$ and theoretical calculations based on Rayleigh-Plesset equation (RPe) \cite{Amans:2016:APL}
   used in the paper \cite{Amans:2016:APL} to define current internal pressure inside a bubble $p(t)$ from $R(t)$
     supposing that ambient pressure in RPe equals to the pressure inside a container with liquid before laser action; here $R(t)$ is a bubble radius.
 We see from Fig. \ref{ris:01-hierarchy} that these data refer to the late stages of relaxation of a system after laser impact.
 This is the main direction of the LAL research for today.


 At the opposite direction of studies of LAL the authors starts namely from ablation in liquid
  \cite{Bulgakov:2017,Povarnitsyn-Itina:LAL:2013,povarnitsyn:ITINA:LAL:2014,LZ-bulk-LAL:2017,LZ+Stephan:2018.LAL,INA.arxiv:2018.LAL,INA.jetp:2018.LAL},
    that is from the beginning of the chain of processes shown in Fig. \ref{ris:01-hierarchy}.
 This is rather new direction of the LAL oriented researches.
 The listed papers describe initial ablation theoretically and by computer simulations.
 The paper \cite{Bulgakov:2017} is exception.
 In this paper the theory is supported with valuable experiments related to the early stage.
 Authors of \cite{Bulgakov:2017} measure reflectivity during a nanosecond laser pulse.
 A temporal range covering a few nanoseconds have been studied in these works.
 In paper \cite{INA.jetp:2018.LAL} simulations have been extended up to submicrosecond stage (up to 0.2 $\mu$s).
 At the sub$\mu$s stage the contact pressure $p_{CB}$ decreases down to critical pressure for water $p_{cr}|_{wt}=220$ bars, see Fig. \ref{ris:01-hierarchy}.


 We consider systems made from the pairs gold-water (Au-wt) and gold-silica (Au-gl, glass).
 Critical parameters are: $T=7.8$ K, $\rho=5.3$ g/cm$\!^3,$ $p=5300$ bar for gold
  \cite{Bushman:1993,Khishchenko2002,lomonosov_2007,rusbank1,rusbank2} (see also Fig. 4 in \cite{INA.jetp:2018.LAL} where saturation pressure of gold is plotted)
   and $T=647$ K, $\rho=0.3068$ g/cm$\!^3,$ $p=220$ bar for water.
 The signs (marking achievement of critical parameters) at the logarithmic axis of time in Fig. \ref{ris:01-hierarchy} corresponds to simulations presented in \cite{INA.jetp:2018.LAL}
     where duration of a laser pulse was $\tau_L=0.1$ ps.
 In the present paper we use data from \cite{INA.jetp:2018.LAL} together with new information obtained in described below simulations corresponding to durations 50 ps and 0.5 ns.
 These durations are marked by the red stars in Fig. \ref{ris:01-hierarchy}.


 In the paper \cite{INA.jetp:2018.LAL} the authors at sub$\mu$s stage drop down in their calculations to critical pressure of water in the near contact boundary (CB) layer.
 This means that the system is close to the stage of formation of a bubble.
 In the present paper we construct one-phase (gaseous) and two-phase (liquid-vapor) adiabatic curves of water.
 These adiabatic curves start from a Hugoniot adiabatic curve in water or from a heated states of water.
 We use this approach to consider not only formation of a bubble but to analyze late stages of bubble expansion.
 At late stages the volume of gaseous water filling a bubble increases many orders of magnitude relative to initial volume of a heated contact water layer.
 Thus we make a bridge between the early (ablation) stages and the late (oscillations of a bubble) stages.
 This is the bridge connecting two directions of studies of LAL.


\section{History of heating of an absorber during a laser pulse}


 There is an expressed ridge $t_s$ dividing ablation regimes to short/fast ones $\tau_L\ll t_s$ and to long/slow (subsonic) ones $\tau_L\gg t_s$ according to duration $\tau_L$ of a pulse.
 This ridge is defined by competition between acoustic phenomena and heating (rate and duration of heating).
 The ridge position at the time axis shown in Fig. \ref{ris:01-hierarchy} is $t_s=d_T/c_s.$
 Here $c_s$ is speed of sound in an absorber.
 Of course, the speed $c_s$ changes during heating.
 For estimates people usually takes a value corresponding to a solid condensed state at room temperature before a laser action; e.g., for gold $c_s=3.1$ km/s.
 For thickness $d_T$ it is reasonable to take thickness of a heat affected zone (HAZ) achieved to the instant $t=t_s.$


 There is a subregime in the short/fast regimes. It is connected to an ultrashort laser pulses (UsLP) when duration $\tau_L$ becomes shorter than equilibration time $t_{eq}$
   mentioned in Fig. \ref{ris:01-hierarchy}.
 Often the corresponding pulses are called femtosecond (fs pulses).
 This name in reality means that the pulse is shorter than picosecond, e.g., 40, 100, or 300 fs.
 Fs pulses are related to the separate class of regimes because their durations are shorter than duration $t_{eq}$ of the two-temperature (2T) stage;
   at this stage electrons absorbing laser energy are much hotter than ions $T_e\gg T_i.$
 Special physics is related to the 2T stage.
 Main features of this physics were revealed in the pioneer paper \cite{Anisimov:1974}.
 Importance of electron-ion coupling characterized by the value $\alpha$ in condensed media becomes clear with this paper.
 Another significant thing is: strong enhancement of electron heat conduction $\kappa$ during a 2T stage
  \cite{Petrov.INA.KPM.jetp.lett:2013,Migdal:spie:2013}
  because semi-degenerate electrons with rather small heat capacity $c$
   are partially decoupled from a classical ion subsystem with large heat capacity \cite{INA:chi:1T:2T:ContribPlPh:2011},
     thus electron thermal diffusion coefficient $\chi=\kappa/c$ is enlarged 10-100 times
  \cite{Petrov.INA.KPM.jetp.lett:2013,Migdal:spie:2013,InogamovSPIE:2013}
     above its usual value $\sim 1$ cm$\!^2/$s corresponding to the one-temperature (1T) conditions;
      degree and duration of decoupling between ion and electron subsystems are defined by the value $\alpha$ - larger $\alpha$ than shorter is equilibration time $t_{eq}$
        defining when the electron-ion temperature relaxation will finishes.



 Electron thermal conduction wave is unusual at the 2T stage.
 This wave carries out from a skin-layer the thermal energy accumulated in electrons \cite{Ashitkov-Au-bulk-eps:2016J.Ph.Conf.Ser};
   the wave is expanding with supersonic speed \cite{Meyer-ter-Vehn:2000}.
 Supersonic expansion of the thermal wave continues during a 2T stage \cite{Petrov.INA.KPM.jetp.lett:2013,Migdal:spie:2013,InogamovSPIE:2013};
  therefore in the case of UsLP (that is when $\tau_L<t_{eq})$ the equilibration time $t_{eq}$ serves as effective duration of a pulse.
 The thermal wave is fast thanks to high values of the electron thermal diffusion coefficient $\chi.$
 Supersonic expansion of heat means that during its existence the spatial expansion of matter is delayed (relative to propagation of heat), that is density remains approximately equal
   to its values before the laser action (isochoric regime); except the thin layer near the contact boundary (CB).
 Thus in the case of a UsLP a HAZ is mainly created during a 2T stage: $d_T=2\sqrt{\chi\, t_{eq}},$
 $$
 d_T(\chi_{1T}=1  \, {\rm cm^2/s}, \, t_{eq}=1 \, {\rm ps})=20 \, {\rm nm}, \;\;\;\;
 d_T(\chi_{2T}=20 \, {\rm cm^2/s}, \, t_{eq}=7 \, {\rm ps})=240 \,{\rm nm}.
 $$
 These estimates emphasize difference between 1T and enhanced diffusions $\chi$ and emphasize the role of the prolonged 2T relaxation (large values of $t_{eq}).$
 For gold the last values (in the above line for $d_T)$ of $\chi_{2T}$ (enhanced electron diffusion) and $t_{eq}$ (extremely delayed relaxation thanks to a heavy ion) are typical
   \cite{Petrov.INA.KPM.jetp.lett:2013,Migdal:spie:2013,Ashitkov-Au-bulk-eps:2016J.Ph.Conf.Ser,Chen_Mo_Soulard_Recoules_Hering_Tsui_Glenzer_Ng_2018}.
 Position of the mark $t_{eq}$ in Fig. \ref{ris:01-hierarchy} corresponds to the case of gold.
 We say -extremely delayed- in the meaning of comparison of gold (Au) with poorly conducting metals like Ni, Pt, or Ta were diffusion is weaker while an electron-ion coupling is stronger
   \cite{Petrov2013b,}.
 We say that the supersonic thermal wave is unusual because we have got into the habit that conductive propagation of heat is very subsonic.


 Above we have underlined the peculiarities of UsLPs $\tau_L\ll t_s$ linked to the 2T states.
 But the main designation of the Section is to separate on a physical ground the short/fast laser actions $\tau_L\ll t_s$ against the long/slow (subsonic) actions $\tau_L\gg t_s.$
 The actions $\tau_L\ll t_s$ is called also the stress confinement actions while the case $\tau_L\gg t_s$ is called the heat confinement regimes \cite{LZ-J-Chem-Ph-C-obzor:2009}.
 There are simple meanings of these terms.
 If $\tau_L\ll t_s$ then during the fast energy absorption the matter remains approximately motionless (isochoric absorption)
   then the largest pressure to absorbed energy ratios $p/F_{abs}\sim 1/d_T$ at the end of a pulse are achieved.
 Then the stress confinement during the time interval shorter than $t_s$ takes place; the pressure bump created by UsLP in a HAZ cannot unload faster.


 After that time interval $t<t_s$ the acoustic decay of the stress confinement proceeds.
 The HAZ irradiates acoustic perturbations in the directions outside from the HAZ during this decay while the heated layer (the HAZ) remains attached to the material particles
    where absorption (skin) and conductive expansion of heat occurred.  
 In the next Section we will consider separation of acoustic and entropy modes and their existence long after the separation that is at the stages $t\gg t_s.$
 While here the acoustic near-field region $t\sim t_s$ is discussed.
 There are acoustic and entropy-vortical modes in hydrodynamics.
 In one-dimensional (1D) case vortex is impossible therefore acoustic and entropy modes are present in adiabatic (without thermal conductivity) hydrodynamics.
 Acoustic modes propagates with sound speed relative to material particles while the entropy $s$ is motionless (it is attached to) relative to these particles.
 If we include thermal conduction then the entropy mode will spread (entropy $s$ will spread) relative to the material particles.



 In the long/slow regime $\tau_L\gg t_s$ the pressures are relatively low; the ratio $p/F_{abs}$ is significantly less than $1/d_T.$
 Separation in time to the near-field temporal region and to the far-field temporal region loses its sense in the case $\tau_L\gg t_s.$
 In the case $\tau_L\ll t_s$ we have a short laser pump and after that the acoustic decay of the HAZ created by the pump.
 While if $\tau_L\gg t_s$ then the heated region (skin and conduction layer) continuously irradiates outside the weak acoustic signals during duration $\tau_L\gg t_s$ of a long pulse.
 The irradiation finishes together with a laser pulse.
 After that the irradiated two trains of acoustic waves propagate to the gold side and to the water side.


 In both $(\tau_L\ll t_s$ and $\tau_L\gg t_s)$ cases effectively motion is a result of thermal expansion of matter.
 The shift of a contact boundary (CB) $\Delta x$ is proportional to a coefficient of thermal expansion $\beta\sim 0.1/T_{cr},$ increase of temperature $T,$ and length $d_T$ of a HAZ:
   $\Delta x \sim (\rho_0/\rho_{HAZ} - 1 ) d_T,$ where $\beta$ is a value of the thermal expansion coefficient at room temperature,
     $\rho_0$ and $\rho_{HAZ}$ are initial density and density after thermal expansion.
 Density $\rho_{HAZ}$ is twice and more times less than $\rho_0$ in the considered here conditions.
 Thus typically the shift $\Delta x$ is a fraction of one micron because $d_T$ is one or few hundreds of nanometers.


 Resuming the last two paragraphs we can say that the shocks sent by the short/fast actions $\tau_L\ll t_s$ have the shorter shock affected layer behind the shock
  (thickness $\sim d_T)$
   and have larger amplitude
    relative to the long/slow actions $\tau_L\gg t_s.$
 While the evolutions of the near CB hot layers after finishing of a pulse differ less; these layers are called also the layers connected to the entropy or advection mode.
 Of course, it is known about breaking of compression waves and the late time fate of the shocks with decrease of their amplitude and widening of compressed layer behind
    due to non-linear acoustic effects
                                 \cite{Zhakhovsky:elastic-plastic:PRL:2011,elastic-plastic:JETPLett:2011,Demaske:2013};
                                   in the case $\tau_L\ll t_s$ dispersion of speed of sound (speed depends on pressure at the characteristics) slowly increases thickness
                                     of the shocked layer behind a shock above the initial thickness $\sim d_T.$
 Also elastic-plastic transition
\cite{Zhakhovsky:elastic-plastic:PRL:2011,elastic-plastic:JETPLett:2011,Demaske:2013,Agranat:JETPLett:may2010,elastic-plastic:JETPLett:2010,Ashitkov:elastic-plastic:JETPLett:2010,Armstrong:elastic-plastic:PRL:2011,elastic-plastic:JPhysConfSer:2014,Perriot:elastic-plastic:JPhysConfSer:2014},
  solid-solid phase transitions \cite{alpha-eps:Fe:Ashitkov:2017,Zhakhovsky:MD:alpha-eps-Fe:AIP:2017},
   and effects of melting \cite{Zhakhovsky:melting-in-SW:PRL:2012}
    are significant for the profile of the shock running inside a solid target.



 Below we will compare the full hydrodynamic simulation (including conductivities of gold and water) and simulation that exclude motion, thus this is the simulation without hydrodynamics.
 In the simulations without hydrodynamics only the thermal problem of absorption of a laser pulse and conductive spreading of heat is solved.
 Sometimes authors limit themselves with this motionless approach saying that motion is very slow and therefore insignificant in the case of a long pulse.
 Hence they neglect variations of density.
 But as was said before, the thermal expansion leads to strong decrease of density in the both cases (fast and slow) if absorbed energy is large enough.
 In the slow case these changes develop slowly but at the end of a pulse they are of the same order as in the case of a fast action.

\begin{figure}       
   \centering   \includegraphics[width=0.8\columnwidth]{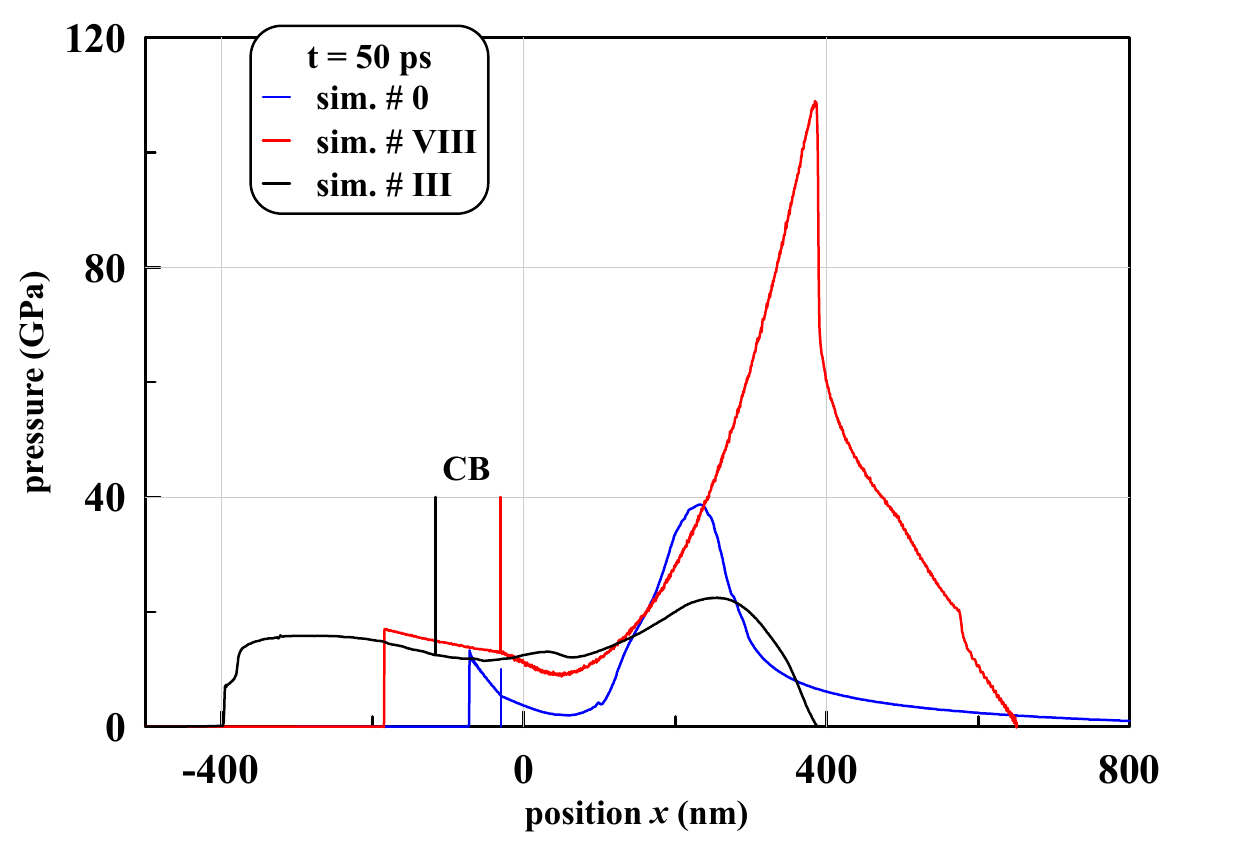}
\caption{\label{ris:02} Comparison of the instant pressure profiles for the simulations listed in the Table 1.
In the case \#0 the total pressure $p=p_e+p_i$ profile is presented. The 2T effects become weak at the shown instant.
The blue, red, and black straight lines mark current positions of the CB in the corresponding simulations
 \#\#0, VIII, and III.
    }  \end{figure}

\begin{figure}       
   \centering   \includegraphics[width=0.8\columnwidth]{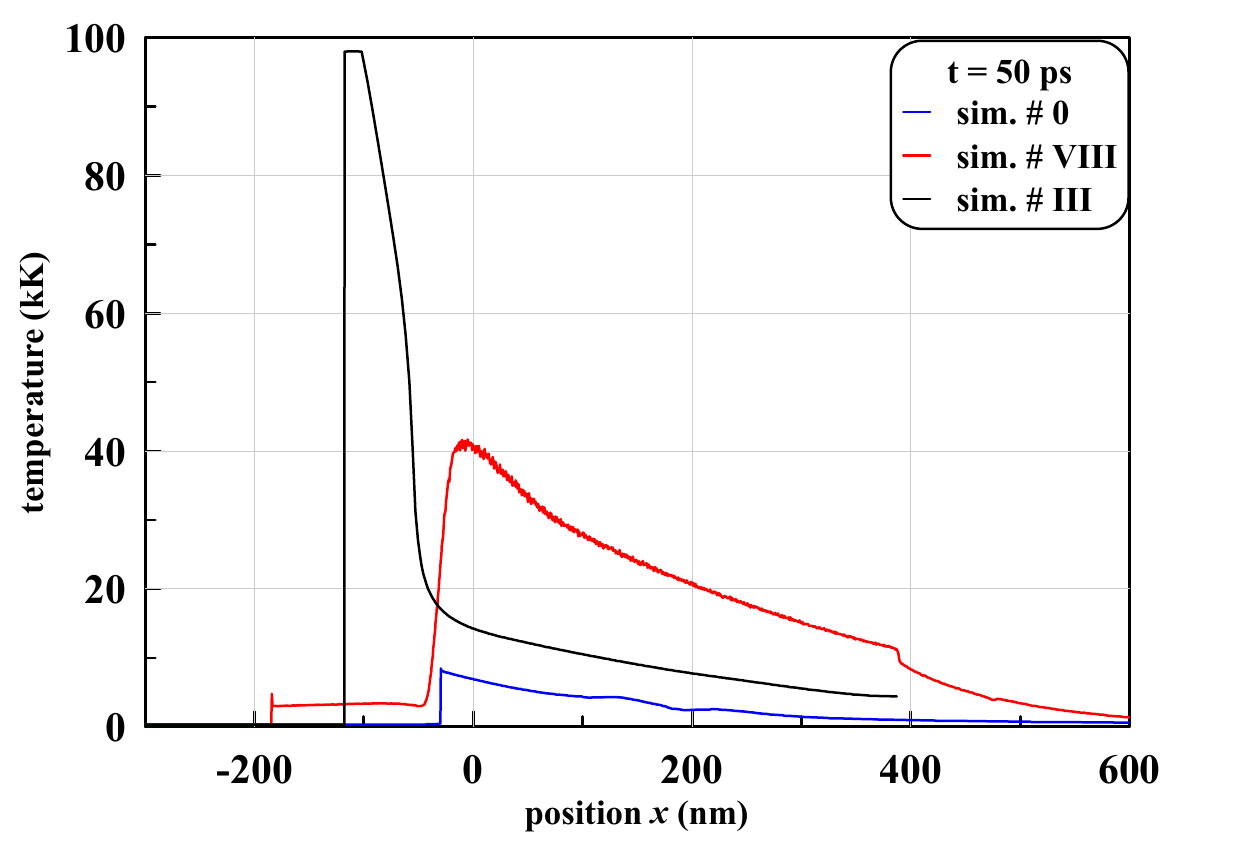}
\caption{\label{ris:03} The temperature profiles for the simulations \#0, III, and VIII listed in the Table 1.
In the case \#0 the ion temperature $T_i$ profile is given. The 2T effects become weak at the shown instant.
    }  \end{figure}


 To better understand and emphasize importance of pulse duration we perform simulations relating to the edges of the fast-slow range of durations and to the middle of this range.
 For gold the $d_T$ is $\sim 150$ nm for fast and slow regimes. Then the acoustic scale $t_s=d_T/c_s$ shown in Fig. \ref{ris:01-hierarchy} is $\sim 50$ ps.
 The ratios $\tau_L/t_s$ are 0.002, 1, and 10 for the fast, intermediate, and slow regimes of laser action presented in Fig. \ref{ris:01-hierarchy} by the three red stars;
    $\tau_L$ is 0.1, 50, and 500 ps for these stars.
 The set of simulations covering the fast/slow range is described in Table 1.
 They are done by 2T-HD, 1T-HD, 1T-without HD, and MD codes; here HD and MD stand for hydrodynamics and molecular dynamics;
  "without HD" means that only thermal equation is solved in the conditions of the isochoric heat absorption and transfer.



 \vspace{0.2cm}
 \noindent
 Table 1. Eight runs with different durations of a heating pulse. Here pairs Au-water or Au-glass are presented.
 Length of simulation in time is given in the column $t_{end}.$

\begin{tabular}{|c|c|c|c|c|c|}
\hline
  \#                           & $\tau_L$ ps & $F_{abs}$ mJ/cm$\!^2$ & medium & $t_{end}$ ns & code       \\
\hline
  0 \cite{INA.jetp:2018.LAL}   & 0.1         & 400                   & water  & 200          & 2T-HD               \\
\hline
  I                            & 50          & 338                   & glass  & 0.5          & 1T-HD               \\
\hline
 II                            & 50          & 559                   & glass  & 0.5          & 1T-HD                         \\
\hline
 III                           & 50          & 897                   & glass  & 0.5          & 1T-HD                        \\
\hline
 IV                            & 50          & 400                   & water  & 1            & 1T-HD                       \\
\hline
  V                            & 500         & 900                   & water  & 1            & 1T-HD              \\
\hline
 VI                            & 500         & 900                   & water  & 10           & 1T-without HD         \\
\hline
 VII \cite{INA.jetp:2018.LAL}  & 1           & 700                   & water  & 1            & MD          \\
\hline
 VIII, wide                    & 1           & 2500                  & water  & 2            & MD       \\
\hline
 IX, narrow                    & 1           & 2500                  & water  & 10           & MD       \\
\hline
\end{tabular}

\vspace{0.2cm}


 Figures \ref{ris:02} and \ref{ris:03} present situation when mixed inside the HAZ the acoustic and entropy modes
   begin to decouple as $t=t_s.$
 Simulations \#0 and VIII correspond to UsLP. But absorbed energy in case \#VIII is approximately five times larger
   than in the case \#0.
 Thicknesses of the gold films are 1700 nm (\#0), 550 nm (\#VIII), and 370 nm (\#III).


 In the case of the UsLP and a thick (bulk) gold target there is a rather strong difference between pressures $p$ in gold and in transparent dielectric,
   see Fig. \ref{ris:02};
    here we compare pressure inside the bulk gold and at a shock front in dielectric;
     pressures in gold and dielectric are equal at the CB.
 Large difference in pressures $p$ is explained by the model with decay of a pressure jump created by an ultrafast heating;
    the jump in this model is linked to the CB \cite{INA.jetp:2018.LAL}.
 Pressures in the contacting materials after decay of a jump (outside the vicinity of the CB)
   depend on their acoustic impedances, see explanations in \cite{INA.jetp:2018.LAL}.
 In the intermediate case $\tau_L\sim t_s$ and in the case with a long pulse $\tau_L\gg t_s$ the pressures profiles
   cannot be explained using the simple model with decay of a pressure jump.




 The shock in transparent dielectric appears almost immediately with the UsLP.
 This is so because the UsLP is sharp, and the pressures (which are created in our two dielectrics, water and glass)
   are larger than a bulk modulus in these dielectrics;
     what is said relates to rather significant energies $F_{abs}$ corresponding to regimes of NPs production.
 This means that the shocks in our cases in the considered dielectrics are strong.
 At the same time the compression wave in gold is weakly or moderately non-linear because pressures in gold
   are less than a bulk modulus 180 GPa of gold.
 Therefore some time proceeds before the compression wave in gold will overturn with formation of a shock \cite{elastic-plastic:JETPLett:2011,Demaske:2013}.
 While in the case with a long pulse the pressures are smaller and increase slowly during a pulse.
 Hence longer time is necessary to wait for their overturning.


 Increasing $F_{abs}$ we increase temperature and decrease density of gold near the CB.
 Also intensity $I_{inc} = F_{abs}/A/\tau_L$ of incident light increases; here $A=1-R$ is an absorption coefficient.
 There is a transition to plasma corona like expansion into vacuum at high intensities $I_{abs}.$
 In a plasma corona the density drops to values much lower than solid state density (orders of magnitude lower).
 In corona the absorption takes place at plasma critical density $\rho_{pl}$ where frequency of laser light equals to local plasma frequency of a rarefied ionized metal.
 In the case with transparent dielectrics two limitations appear at the way to corona like expansion.

\begin{figure}       
   \centering   \includegraphics[width=0.8\columnwidth]{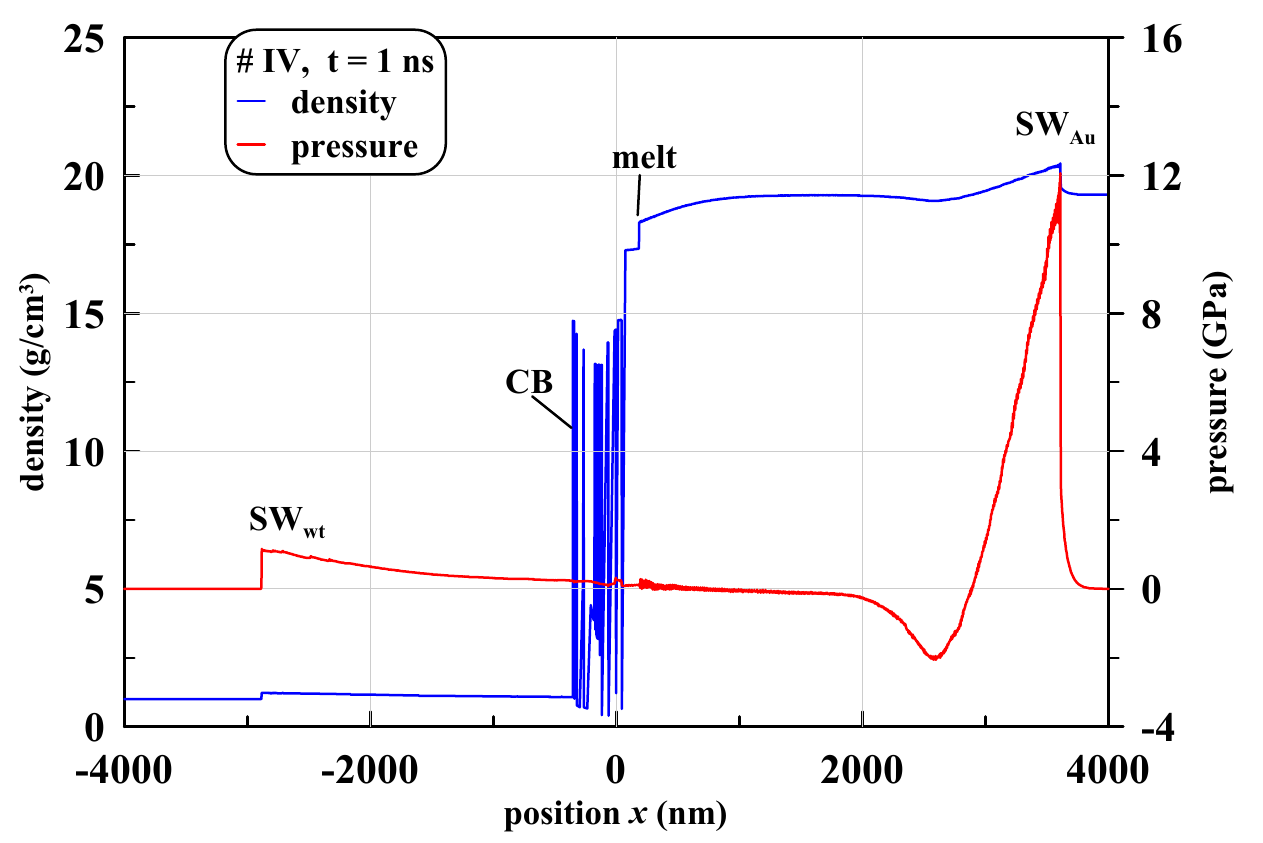}
\caption{\label{ris:04} Spatial separation of acoustic signals (they are the triangular shock waves (SW))
 and the entropy mode - the heated layer near the CB (contact boundary).
    }  \end{figure}

\begin{figure}       
   \centering   \includegraphics[width=0.8\columnwidth]{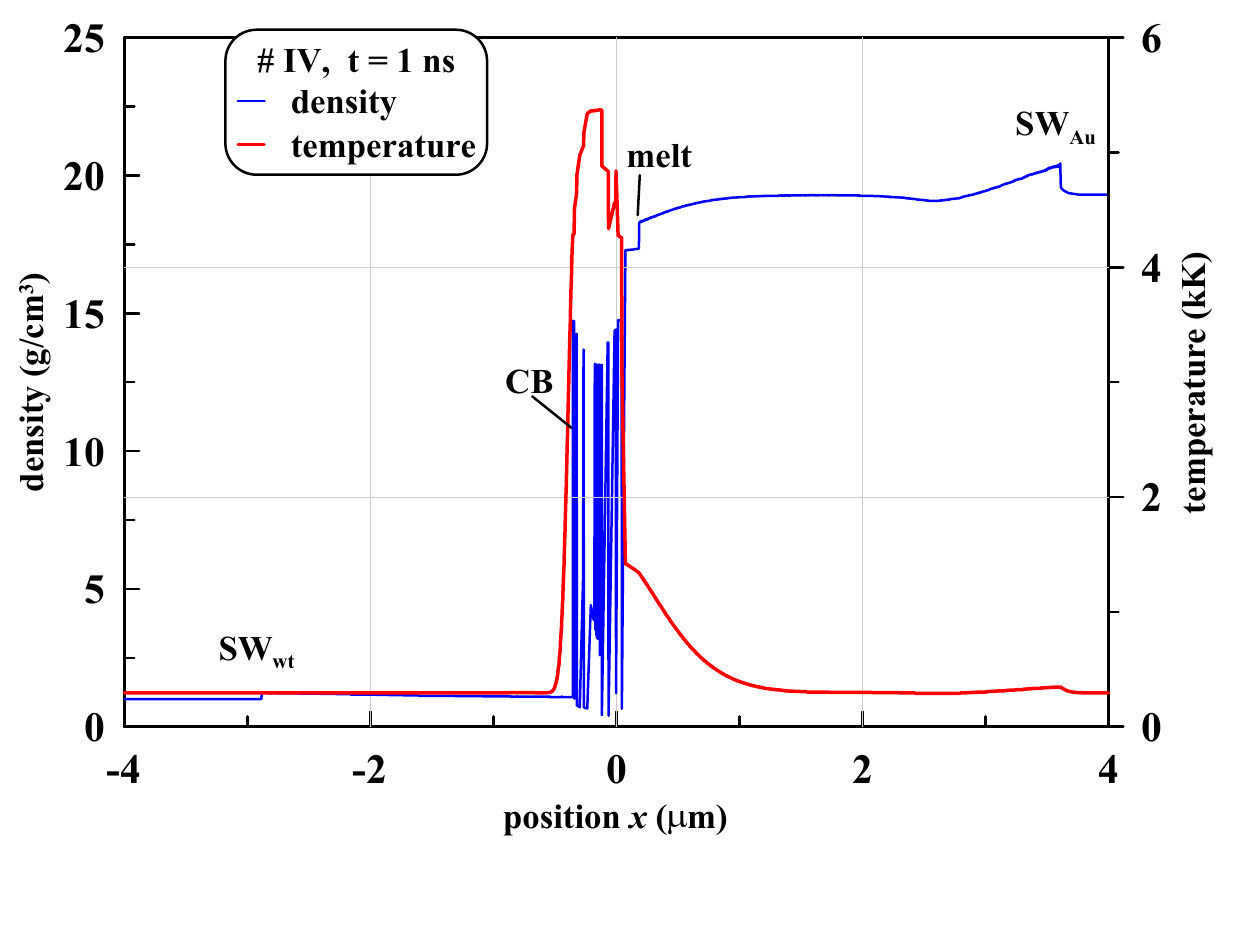}
\caption{\label{ris:05} Temperature profile corresponding to the same instant $t=1$ ns as shown in Fig. \ref{ris:04}.
 We see that SWs weakly heat matter.
 We should separate heating by reversible (adiabatic) heating thanks to compression in the triangular region behind a SW
   and dissipative heating in a SW which remains in matter after passage of a SW.
 Thermal energy remains concentrated inside the entropy mode called also advection layer
   because it doesn't move relative material particles with speed of sound as an acoustic mode
    but stays frozen into matter and moves (that is advected) together with matter.
 Entropy mode locates where it was introduced into matter, in a skin-layer.
 It spreads out from the skin thanks to thermal conductivity of matter.
 Conductivity of gold is two-orders of magnitudes higher than that in water,
   therefore the thermal spread into the gold side is much wider.
 At the instant shown, the front "melt" marks not a melting but a recrystallization front, it moves to the left side with velocity 180 m/s, see explanations in the text.
    }  \end{figure}





 First, we cannot irradiate metal with very large intensities because an optical breakdown will shadow metal surface;
   may be the decrease of reflectivity observed in \cite{Bulgakov:2017} during a laser pulse corresponds to the pre breakdown phenomena?
 Second, density of expanding metal cannot be so easily decreased below plasma critical density $\rho_{pl}$
   during the nanosecond pulse
    due to confinement
     by dense dielectric; situation is very different from expansion to vacuum;
      for gold $\rho_{pl}\sim 0.1$ g/cm$\!^3$ for optical lasers and ionization degree $\sim 1$ (one ionized electron per atom).
 We see examples with high temperatures (and hence low densities) in Fig. \ref{ris:03}.
 But still they are far from plasma critical density $\rho_{pl}$ for corona.
 Corresponding absorbed fluences in our simulations (see Table 1) relate to the edge of the largest of fluences used in typical experiments
   \cite{Stephan:2017.review,XIAO:2017}.

\begin{figure}       
   \centering   \includegraphics[width=0.8\columnwidth]{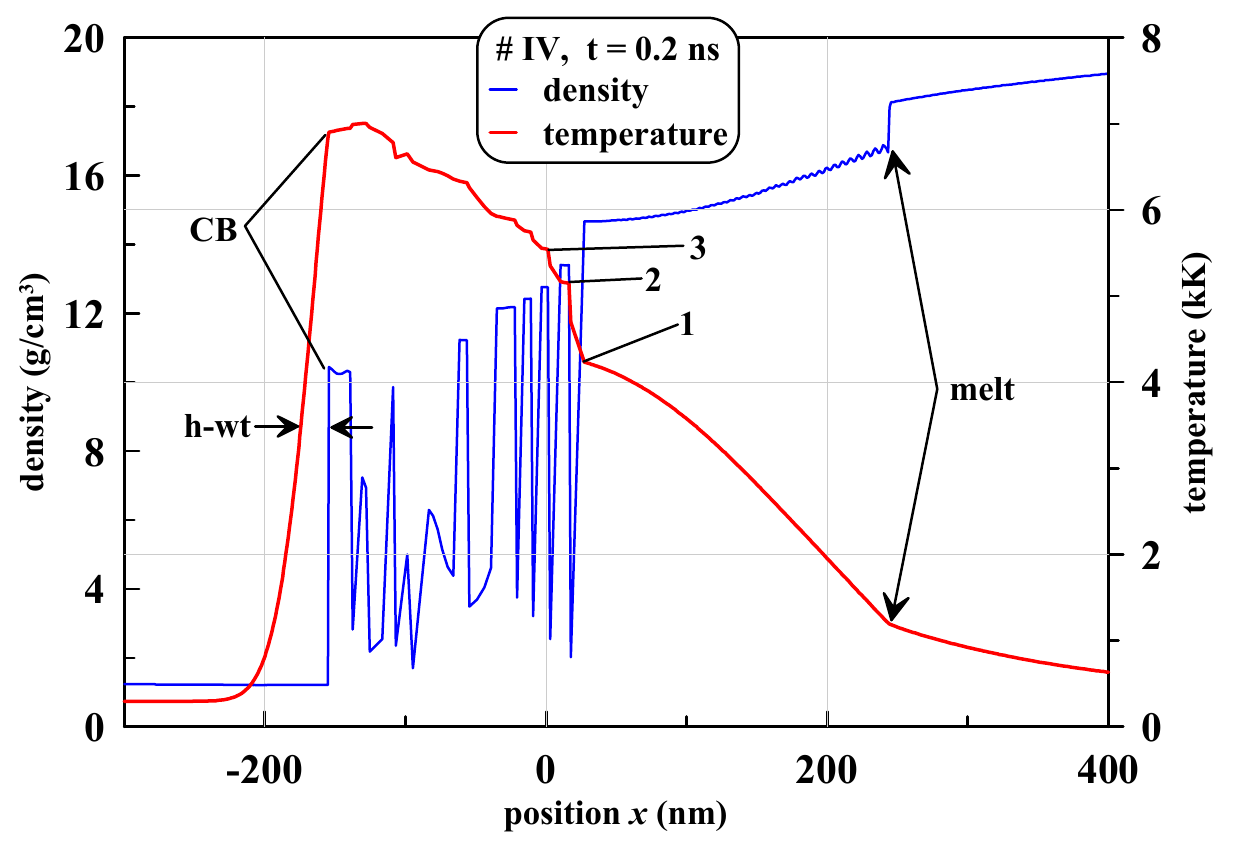}
\caption{\label{ris:06} Density and temperature profiles.
The note "melt" marks current position of the melting front.
Melting causes decrease of density. The molten phase is to the left side relative to the front.
Thermal conductivity $\kappa$ drops approximately 2.5 times after melting.
Thanks to that and to energy expenses to melt gold the gradient of temperature is larger at the liquid side.
Conductivity spreads thermal energy from a HAZ to the right side.
This energy flux melts solid.
The density jump at the stage shown is indeed a melting front moving to the right (to the solid side)
  at the instant shown with velocity 350 m/s relative to material particles.
Hydrodynamic velocities are 70 and 100 m/s at the melting front at the its solid and liquid sides, respectively.
These velocities are directed to the left side (to the side of water).
    }  \end{figure}

\section{Separation of acoustic zones and hot advection layer}




 Let's consider the era $t\gg t_s$ that comes after decoupling of acoustic and entropy modes in a HAZ.
 This era is definitely seen in the case of short/fast regimes $\tau_L\ll t_s.$
 While in the case of a long pulse this era begins later $t\gg \tau_L > t_s.$
 Fig. \ref{ris:04} demonstrates the typical situation relating to the era; here the ratio $t/t_s=20,$ $t=1$ ns, $t_s\sim 50$ ps, $\tau_L=50$ ps.
 Later in the era the situation doesn't change in the sense that the triangular SWs and the advection layer are far away
   and become almost independent from each other.
 At the late stages of the era, it is difficult to plot the SWs and advection layer together
   because their thicknesses become very small relative to the spaces which separate them.


 Fig. \ref{ris:05} proves that heat is kept inside the entropy mode located
   where the primary absorption of laser energy takes place - around the skin-layer.
 In the case of optical lasers the skin is a thin (10-20 nm thick) layer at the CB.
 Heat conductivity smears thermal energy distribution spreading heat from a skin.
 Spreading is very subsonic thus a thin clearly seen front of subsonic melting appears in Fig. \ref{ris:05};
  this is a density jump marked as "melt" in Fig. \ref{ris:05};
   let's mention here that melting at the 2T stage in case of the UsLP
   is smeared covering a significant part of a supersonic thermal wave \cite{INOGAMOV20099712}
    (such type of melting is called homogeneous nucleation of liquid phase; 2T state means UsLP, then $p$ are high,
      then $T_{melt}(p),$ $\rho\approx \rho_0);$
        thus at the 2T stage the Stefan problem approximation of melting/freezing isn't applicable,
          while it is applicable in the situations shown in Figures \ref{ris:05} and \ref{ris:06}.
 In these Figures the melting/freezing front, first, marks itself by a density jump (see note "melt").
 The jump corresponds to the edges of the triple point at the $\rho,T$ or $\rho,p$ phase planes.
 The jump appears due to spatial expansion of liquid phase of gold relative to the solid phase
   at the temperature of melting.
 Second, temperature at the jump equals to temperature in the triple point; 1337 K for gold.


 It isn't necessary to solve separately the Stefan problem in our approach where hydrodynamic equations
  (including energy balance with heat conduction)
    are coupled with equation of state (EoS) of real matter.
 The problem with melting/crystallization is solved inside these equations
  because the EoS describes phase transitions (melting, evaporation)
   and takes into account heat of fusion and heat of vaporization.
 Usually the Stefan problem is solved using only energy balance, heat of fusion, and melting temperature.
 In our approach the hydrodynamics relating to melting is present.
 Thus movement of matter due to expansion of a substance as a result of melting is described; see the caption to Fig. \ref{ris:06}
   where it is written about the difference in hydrodynamic velocities in the solid and liquid sides of the melting front.


 We follow carefully thermal evolution of the entropy mode at the stage (beginning from $t= -150$ ps)
   preceding to the instant 1 ns shown in Fig. \ref{ris:05};
     we start the 1T-HD simulations (with and without HD) at the instant equal to $-3\tau_L.$
 One example with $t=200$ ps is presented in Fig. \ref{ris:06}.
 This Figure describe the situation with heating and formation of foam in more details
   relative to Fig. \ref{ris:05} showing a global structure.
 What keeps temperature high inside the foam?


 In simulation \#IV (see Table 1 where the simulations are listed) temperatures inside the entropy mode were 9.4 kK
   at the end $(t=50$ ps) of absorption of the $\tau_L=50$ ps pulse.
 The full width at half maximum of the temperature profile $T(x,t= 50 \, {\rm ps})$ is 180 nm.
 After that temperature gradually decreases mainly due to heat conduction into bulk gold and adiabatic cooling;
   heat conduction in water is included but it is small and weakly affects temperature distribution in a metal
     at the rather early stages.

\begin{figure}       
   \centering   \includegraphics[width=0.8\columnwidth]{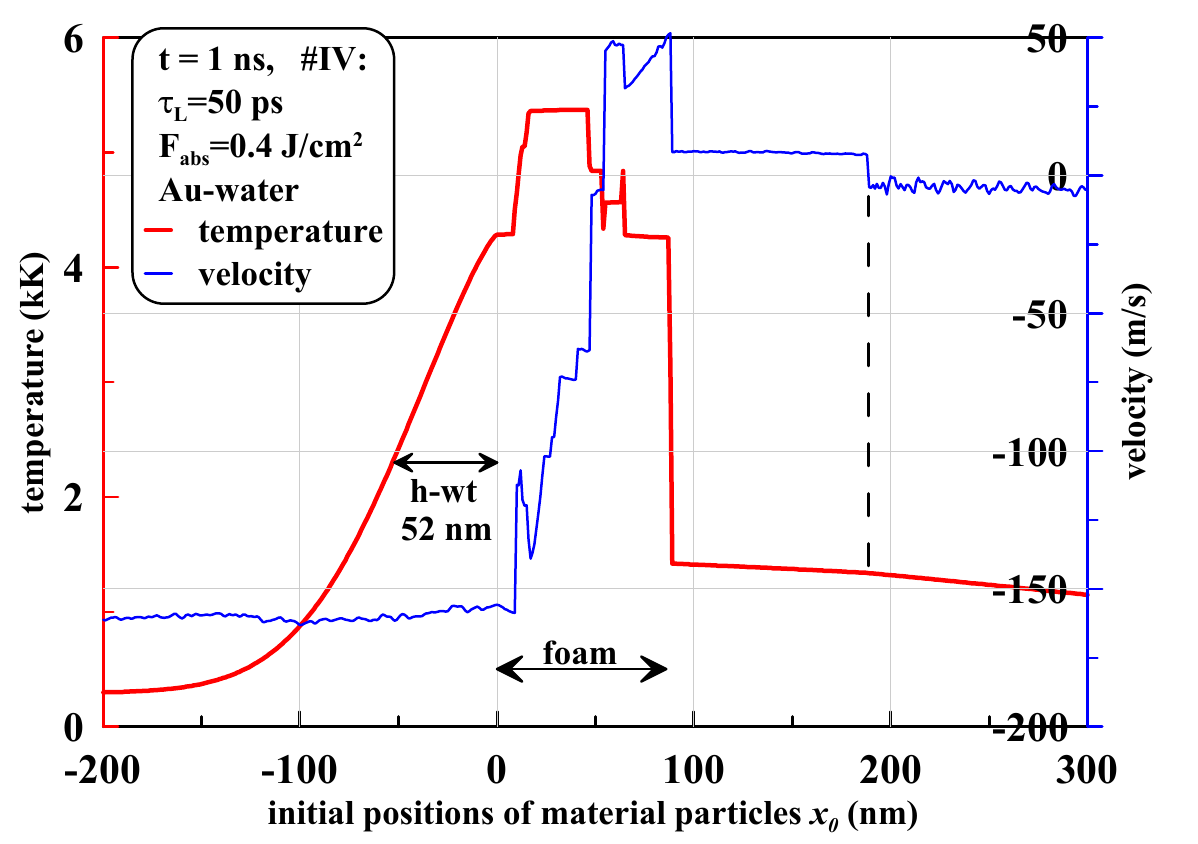}
\caption{\label{ris:07} Internal structure near the contact boundary (CB) marked as "CB" in Figures \ref{ris:05} and \ref{ris:06}.
 At the Lagrangian axis $x_0$ the position of the CB is motionless and equal to $x_0=0,$ gold is to the right side, water to the left.
 The entropy/advection mode consists from foamy layer "foam" and the left warm part of the continuous gold.
 It is reasonable to add the heated layer of water "h-wt" to this mode.
 This layer begins at the hot CB.
 It is also shown in Fig. \ref{ris:06}.
 We did MD simulations to estimate importance of the Kapitza temperature jump effect (Kapitza resistance) and show that in our conditions it is of minor importance, see text.
     }  \end{figure}

 Nucleation of future foam begins at $t\approx 80$ ps in simulation \#IV.
 During few tens of picoseconds the nucleation process covers the layer $d\approx 90$ nm of gold;
  this layer is marked as "foam" in Fig. \ref{ris:07}.
 Mass thickness of the foamy layer is $\sigma=d\,\rho_0 = 1.7\cdot 10^{-4}$ g/cm$\!^2,$ where $\rho_0=19.3$ g/cm$\!^3$ is initial density of gold.
 After that the column mass $\sigma$ doesn't change in time.
 While geometrical thickness $d_{geom}$ of the foamy layer increases with time.
 The $d_{geom}$ is 430 nm at the instant $t=1$ ns shown in Fig. \ref{ris:05}.

 Foaming strongly decreases thermal conductivity of gold.
 This is the answer to the question: what keeps entropy mode hot in Fig. \ref{ris:05}.
 In 1D geometry used in our hydrodynamic codes the foamy zone consists from alternating liquid
  (the digits 2 and 3 in Fig. \ref{ris:06} mark two of these liquid layers) and vapor layers (the digits 1-2 in Fig. \ref{ris:06} mark the first vapor gap from the side of gold).
 Namely the vapor layers weakly conduct heat while thermal conductivity in the liquid layers remains high.
 Therefore the temperature profile in Fig. \ref{ris:06} in the foamy zone is composed from the ladder of steps: the digits 1, 2, 3,... in Fig. \ref{ris:06}.
 The steps are approximately homogeneous temperature distributions inside the liquid layers,
   while the intervals between the steps are the rather steep rises of temperature in a vapor layer
     between two neighboring liquid layers.
 In real 3D geometry the foamy zone contains a mixture of membranes, droplets, and vapor
   \cite{INA.jetp:2018.LAL,Inogamov:2014:subSurfVoids:JPCS,Inogamov:2014ccp,Ashitkov:QuantEl:foam:2014}.
 Vapor surrounding the droplets is weakly conductive.
 Thus droplets lose their thermal contact with bulk gold and hence remain hot.
 Some contact exists along membranes when they are attached to the bottom of a crater.




 As was said, Fig. \ref{ris:06} explains that temperature of a foam decreases slowly with time therefore the bump of temperature is high in Fig. \ref{ris:05}:
   vapor serves as a thermal insulator.
 Continuous gold adjoining on the foam follows its own thermal history in large extent independent from cooling of foam.
 The point of contact between the foam and the continuous gold is denoted by digit 1 in Fig. \ref{ris:06}.
 There is a large temperature jump in this point accumulated in time (comp. Fig. \ref{ris:06} from one side and Figs. \ref{ris:05}, \ref{ris:07} from the other side).
 This accumulation takes place due to low conductance of the isolating vapor layer and cooling by heat transfer into bulk of continuous gold.


 The melting/recrystallization front moves along matter to the right when there is enough internal energy at the left side -
   then the first order phase transition front is the melting front; its velocity to the right at $t=0.2$ ns is given in a caption to Fig. \ref{ris:06}.
 While later in time the thermal reservoir at the left side is exhausted and the melting changes to the recrystallization \cite{INOGAMOV20099712,Volkov:Melt:Solidif:2007}.
 Then the front moves to the left side along material particles.
 In Fig. \ref{ris:05} corresponding to $t=1$ ns there is a recrystallization front moving to the left side with velocity $v_{recr} = -180$ m/s relative to matter.
 Then soon $\approx t=2-3$ ns whole continuous gold adjoining on the foam will be solid.


 Typical maximum velocity of recrystallization of a metal is $\sim 100$ m/s.
 Rate of diffusion of atoms in a liquid phase limits velocity of recrystallization $v_{recr}$ \cite{Wai-Lun:PRB:2008, Inogamov:2014:subSurfVoids:JPCS, Inogamov:2014ccp}.
 Our hydrodynamic simulations gives larger values for this quantity corresponding to formal application of a Fourier's law for thermal conductivity
  connecting $\nabla T$ and $v_{recr}$ \cite{INOGAMOV20099712,Volkov:Melt:Solidif:2007}.
 Of course, a real situation with the subnanosecond pulse isn't so close to equilibrium as it supposed by a Fourier's law, thus it differs from an equilibrium picture
    (but not strongly: velocities $v_{recr}$ are comparable).
 From kinetic description it follows that: (i) a liquid phase is overcooled below the triple point temperature;
   (ii) there is smearing of recrystallization front due to nucleation of nanocrystals in bulk of a liquid ahead of the
   front;
     and (iii) solidified gold transits into polycrystalline state with extremely small sizes (few nanometers) of the crystalline grains
 \cite{LZ-bulk-LAL:2017,LZ-J-Chem-Ph-C-obzor:2009, Wai-Lun:PRB:2008, Ivanov:rost:nnoBump:film:substrate:JAP:2010, Inogamov:2014:subSurfVoids:JPCS, Inogamov:2014ccp, Ivanov:filmSeparFromSubstrate:2013, Inogamov2016APA, Inogamov2016nanoscResLett, AnisimovFLAMN:2017}.


 Foamy part of the entropy/advection mode covers the column mass $0<x_0<90$ nm in Fig. \ref{ris:07}.
 Here and in Fig. \ref{ris:07} the coordinate $x_0$ is Lagrangian coordinate.
 It is equal to initial (before laser action) position of a material particle at the axis $x$ perpendicular to the surface.
 Velocities inside the layers oscillates due to slowly decaying acoustic modes coupled to the liquid layers of the foamy zone.
 The acoustic modes are triggered by the events of rupture of continuous molten gold.
 Where the foam nucleation wave stops when it propagates into bulk of gold depends on strength of gold and tensile stress created due to expansion into medium
   with lower acoustic impedance.
 Strength of gold depends on phase state (solid is stronger) and temperature of liquid.
 Strength decreases as temperature increases.
 Therefore at some depth the nucleation wave stops.
 This depth define thickness of the foamy zone along material axis $x_0$ shown in Fig. \ref{ris:07}, this thickness is underlined by the arrow "foam" in Fig. \ref{ris:07}.


 Solidification is accompanied by contraction of gold.
 This causes appearance of mass flux of liquid to the solidification front.
 Velocity of this flux is $\approx 20$ m/s.
 This value defines the jump of velocity in Fig. \ref{ris:07}.
 Position of the jump is marked by dashed vertical straight line in Fig. \ref{ris:07}.
 Solidification/recrystallization front is shown by intersection of this dashed line with a temperature profile in Fig. \ref{ris:07}.
 Solidification returns back heat of fusion.
 This process slows down the rate of decrease of temperature in the entropy/advection mode outside the foamy zone.


 All simulations \#\#IV-VIII listed in Table 1 include heat conductivity $\kappa_{wt}$ of water;
   in \#\#IV-VI we take $\kappa_{wt}=0.6$ W/K/m and heat capacity 4.2 J/K/g, then thermal diffusivity is $\chi=1.4\cdot 10^{-3}$ cm$\!^2/$s for normal density water.
 In molecular dynamics (MD) simulations \#\#VII,VIII heat conductivity is defined by the used interatomic potential for water, description of this item needs separate discussion.
 Water is heated through the CB from hot gold.
 Using MD simulations we estimate the Kapitza resistance for water-gold interface.
 This type of resistance is linked to difference of mass of atoms and atomic character of heat conduction in water;
   in MD simulations water is described as a point atoms.
 MD simulations show that temperature jump due to Kapitza resistance is of the order of 100 K; presentation of the corresponding calculations is out of the frame of this paper.
 This value is small relative to multi kilo Kelvin temperatures of gold in our conditions.
 Another problem is connected with appearance of foam and decrease of heat conductivity on the side of a gold target.



 Temperature of water near the CB is high, see Figures \ref{ris:06} and \ref{ris:07}.
 Hot thin layer of water appears thanks to thermal conduction.
 It is marked as "h-wt" in Figures \ref{ris:06} and \ref{ris:07}.
 The line "h-wt" is plotted at the half maximum of a temperature profile (initial temperature 300 K is subtracted).
 The estimate $2\sqrt{\chi\, t}$ gives 25 nm for $\chi=1.4\cdot 10^{-3}$ cm$\!^2/$s and $t=1$ ns.
 This is approximately a half of thickness "h-wt" shown in Fig. \ref{ris:07}.


 In the Section 3 we describe (A) gradual spatial separation of acoustic and entropy modes (Figures 4 and 5).
 (B) We see that interplay of momentums directed to the right and to the left sides from the heat affected zone
   causes fragmentation and foam formation
    {\it inside the entropy mode} in case of short pulses (fs and multi-ps); Figures 4-7.
 (C) Thermal conduction in gold and in water spreads heat accumulated in the entropy mode; Figures 6 and 7.
 Low heat conductivity of foam and gaseous gold decreases rate and scale of this spreading;
   see next Section about overcritical and hence gaseous like gold.
 Thus foam and gaseous gold for long time remain hot.


\begin{figure}       
   \centering   \includegraphics[width=0.75\columnwidth]{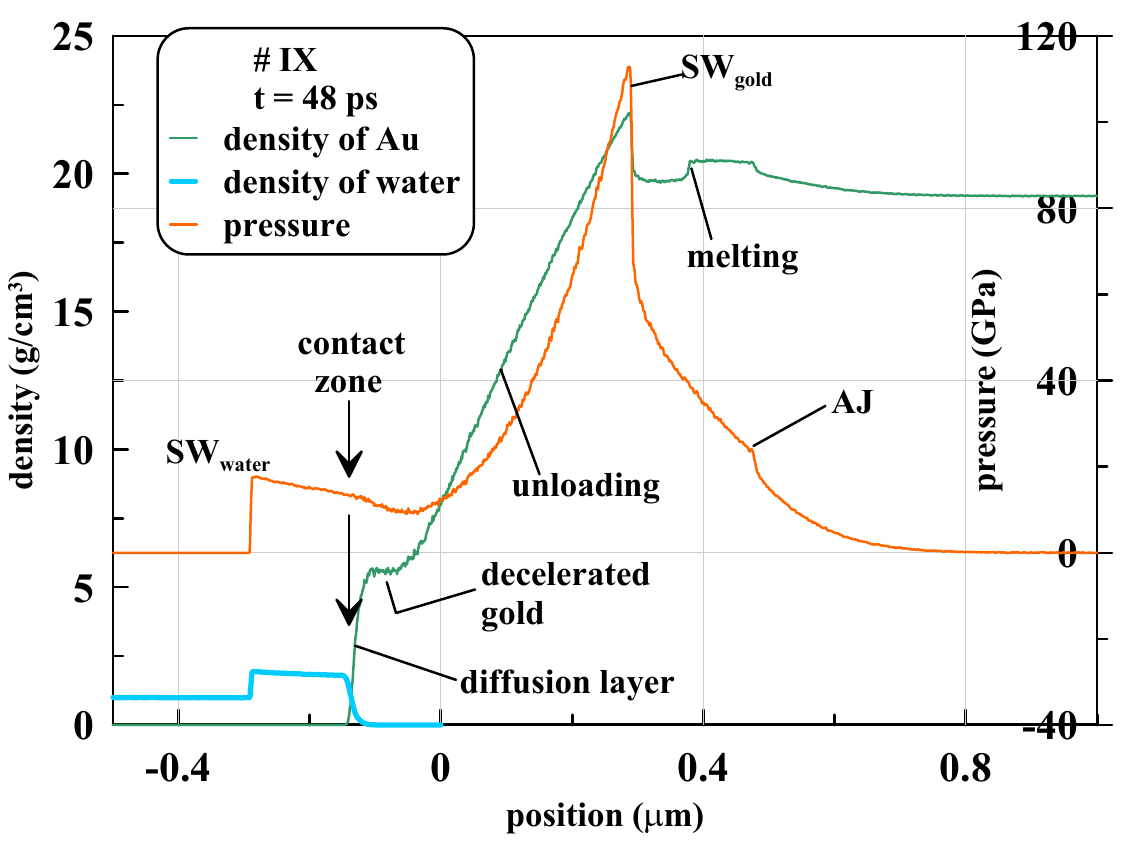}
\caption{\label{ris:08} Structure of flow at the stage when a compression wave leaves the heat affected zone,
 see text for explanations.
     }  \end{figure}

\begin{figure}       
   \centering   \includegraphics[width=0.75\columnwidth]{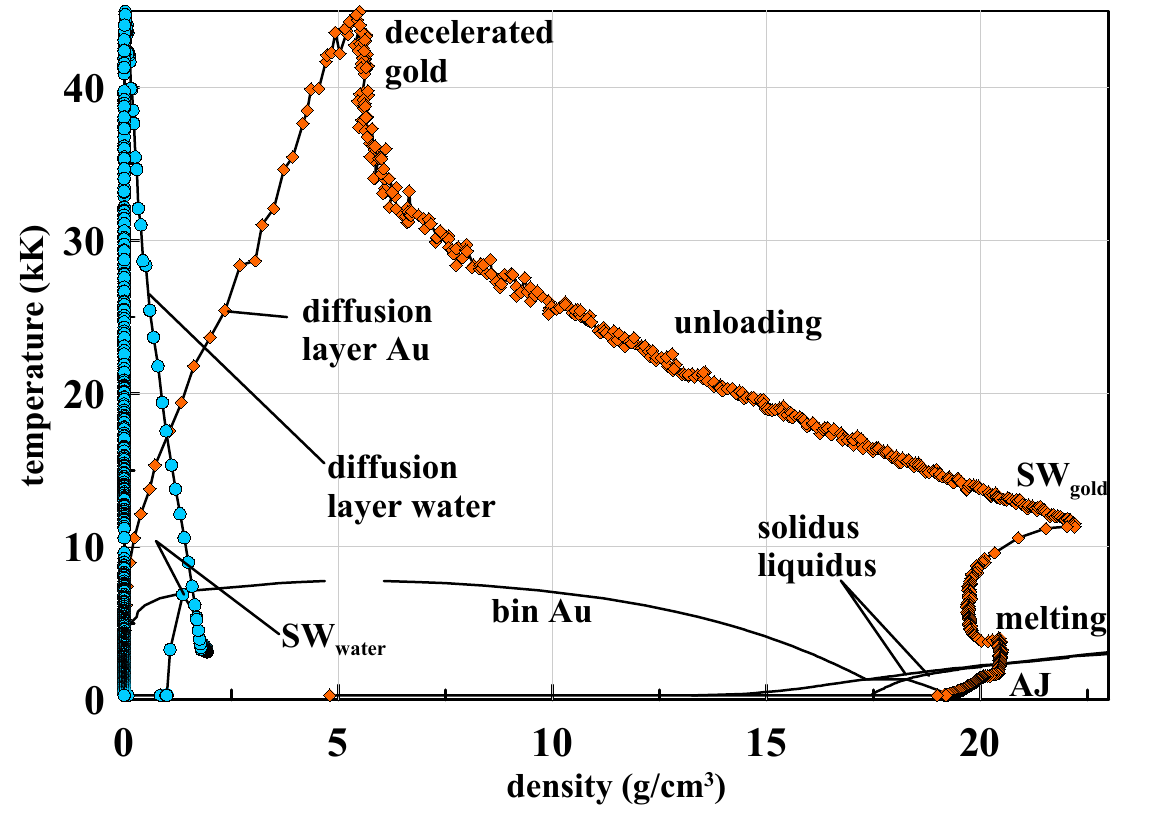}
\caption{\label{ris:09} Binodal ("bin Au") and a melting region between solidus and liquidus are shown
  together with the structure for the instant $t=48$ ps presented in Fig. \ref{ris:08}.
  We see that absorption of 2.5 J/cm$\!^2$ lifts gold into overcritical region.
  The orange rhombuses correspond to gold, the blue circles correspond to water.
     }  \end{figure}



 \section{Overcritical states}



 \subsection{Early stage}


 Absorption of energy $F_{abs}$ of the order of one - few J/cm$\!^2$ transfers gold in its heat affected zone (HAZ)
  into hot dense states above a critical point of gold.
 Let's follow evolution of a gold-water system at later times.
 Fig. \ref{ris:08} presents early structure formed shortly after powerful ultrashort laser impact
   (MD simulation \# IX, see Table 1).
 Structures gradually transform during their dynamic evolution, see below what and how is changed.


 Absorption and thermal spreading of absorbed energy from a skin layer during a two-temperature stage creates a hot layer called HAZ.
 If we forget about light propagation through water and corresponding optical effects connected with refractive index of transparent water,
   because we operate with absorbed energy $F_{abs},$ then we can say that the HAZs in case of water and in case of vacuum are the same as long as the values of $F_{abs}$ are equal.


 Further in time at the acoustic time scale $t_s=d_T/c_s > t_{eq}$ the HAZ formed during the time interval $t<t_{eq}$ corresponding to the two-temperature stage
   acoustically decays into the entropy mode and into two acoustic compression waves running one to the contact with water and the other one to the side of bulk gold;
    let's call the last wave CW-bulk - compression wave propagating to the bulk side; $d_T$ is thickness of a HAZ.
 The entropy mode is the rest of the HAZ after irradiation of acoustic waves from the HAZ.
 The acoustic compression wave running to the contact reflects from the contact as a rarefaction wave (RfW) running behind the CW-bulk.
 Reflection and the reflected wave RfW of course depends on mechanical properties of water -
   in linear acoustics on acoustic impedance of water.


 The instant $t=48$ ps shown in Fig. \ref{ris:08} approximately corresponds to the scale $t_s\approx 50-70$ ps; in gold $d_T\approx 150-200$ nm, $c_s=3.1$ km/s.
 Laser action is strong, thus the CW-bulk quickly overturns forming a shock SW$_{gold}$ (SW - shock wave).
 The RfW sits at the front of the SW$_{gold}.$
 The RfW is denoted as "unloading" (into water) in Fig. \ref{ris:08}.
 The sitting or "attaching" to the front means that in Fig. \ref{ris:08} the SW$_{gold}$ separates the flow to two regions:
   the one at the right side relative to the SW doesn't know about expansion into water, while the other, at the left side,
     depends on mechanical properties of medium surrounding a gold target, i.e. it knows about water.
 The right region is the same as in the case with expansion into vacuum at equal $F_{abs}.$


 Let's stop our discussion about casuality and present other features of the structure shown in Fig. \ref{ris:08}.
 AJ is an acoustic jump. It forms when the CW-bulk transits through the melting front.
 At the two-temperature state and large overheating above melting temperature $T_{melt}(p),$
   the melting zone propagates supersonically; this regime is also called quasi-homogeneous melting.
 A pressure profile created during the time interval $t_{eq}\ll t_s$ at the two-temperature stage
  thanks to approximately isochoric heating
   has more steep spatial piece of the profile in the place where the melting zone locates at the transition stage
    from two-temperature to one-temperature regime.
 Derivative $\partial p/\partial x$ is steeper at this piece because the isochoric pressure rise with temperature
   $\partial p/\partial T|_\rho$ as function of temperature $(\rho$ is fixed)
     is larger at the isochoric melting interval between solidus and liquidus.
 At the acoustic stage this steep piece propagates along sonic characteristics with speed of sound.
 I.e. the two-temperature supersonic melting is imprinted into profile of the CW-bulk.
 Speed of sound depends on pressure.
 Thus gradually the steep piece becomes even more steep.
 At the instant present in Fig. \ref{ris:08} this piece is the AJ.



 In Fig. \ref{ris:08} the melting front is marked.
 At the stage shown the front propagates to the right side increasing mass thickness of molten gold.
 The SW$_{gold}$ and RfW in Fig. \ref{ris:08} were discussed above.


 Gold at the plateau at the left side of the RfW near the contact with water moves approximately with the same velocity
   as the contact.
 This is typical for solutions with acoustic decay of a jump separating two homogeneous semi-spaces.
 A high pressure semi-space produces a shock in a low pressure semi-space.
 While a rarefaction wave propagates into the high pressure semi-space.
 There is a region of homogeneous flow (a plateau) covering a shock compressed layer and a piece of matter
   belonging to the initially high pressure semi-space.
 Stretching of matter is low at the plateau while it is finite in the RfW.


 In our case the jump between semi-spaces separates the initially homogeneous semi-space (water)
   and a high pressure region of finite thickness $d_T.$
 Therefore situation changes relative to classical decay of two homogeneous semi-spaces
   as RfW runs out from the region $d_T.$
 Contact begins to decelerate, density outside the plateau decreases down to density at the plateau.
 Deceleration of the contact causes deceleration of the SW$_{water}$ in Fig. \ref{ris:08}.
 And triangular shape of a water shock and the profile behind it gradually forms.


 Densities of gold and water are shown separately in Fig. \ref{ris:08}.
 We see that in the diffusion zone the concentrations of mixed gold and water gradually changes from 100\% to 0.
 Diffusion smears contact into a mixing zone (arrow "contact" in Fig. \ref{ris:08}).

\begin{figure}       
   \centering   \includegraphics[width=0.75\columnwidth]{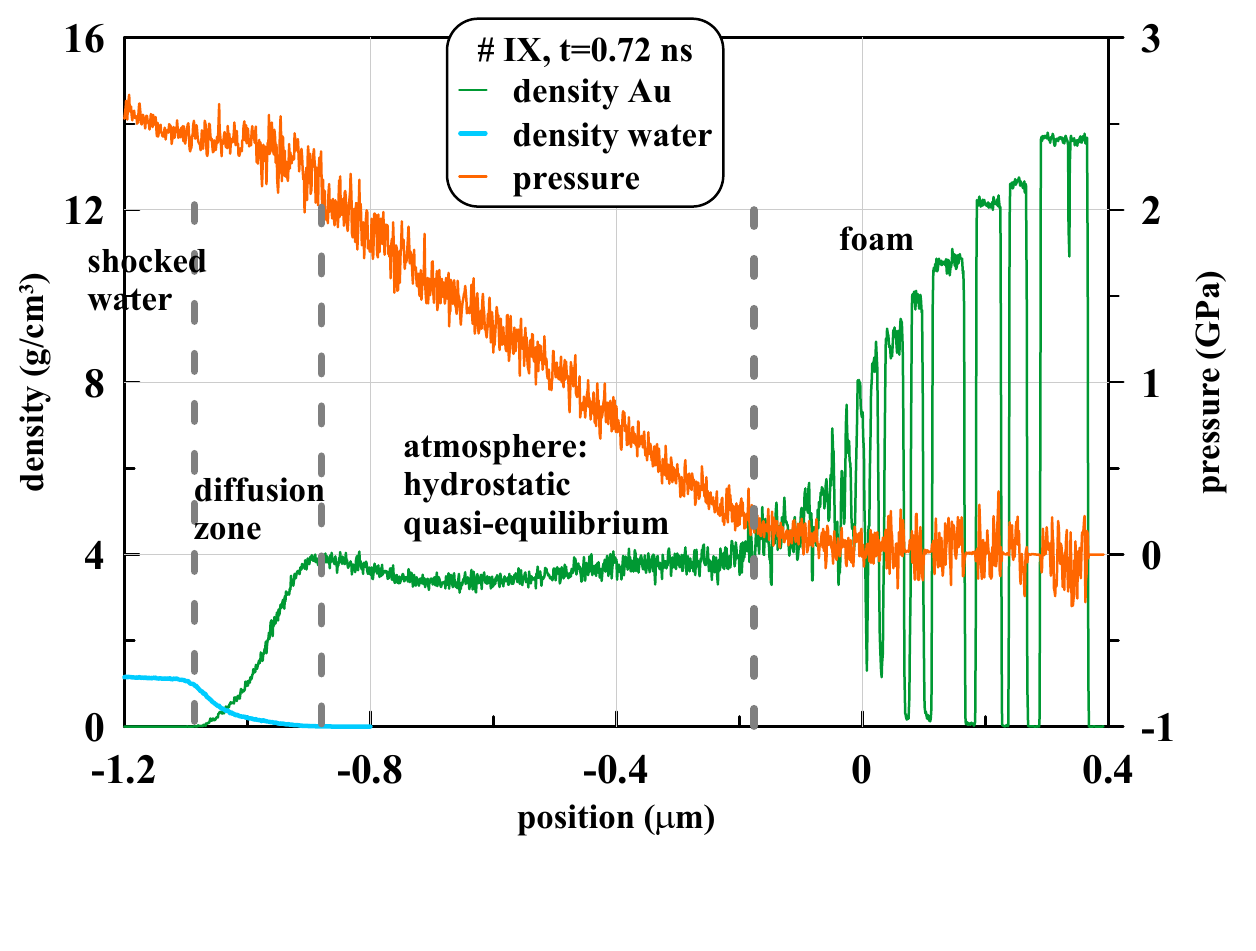}
\caption{\label{ris:10} Pressure and density profiles plotted separately for gold and water are shown.
Structure of flow is described. There are from right to left: foam, atmosphere, diffusion zone, and shock compressed water.
We see that pressure is almost zero in foam.
Gradient of pressure in atmosphere is balanced by effective weight in the non-inertial frame attached to the contact.
Density in atmosphere varies weakly due to rise of temperature of gold in direction to contact, see next Figure.
     }  \end{figure}

\begin{figure}       
   \centering   \includegraphics[width=0.75\columnwidth]{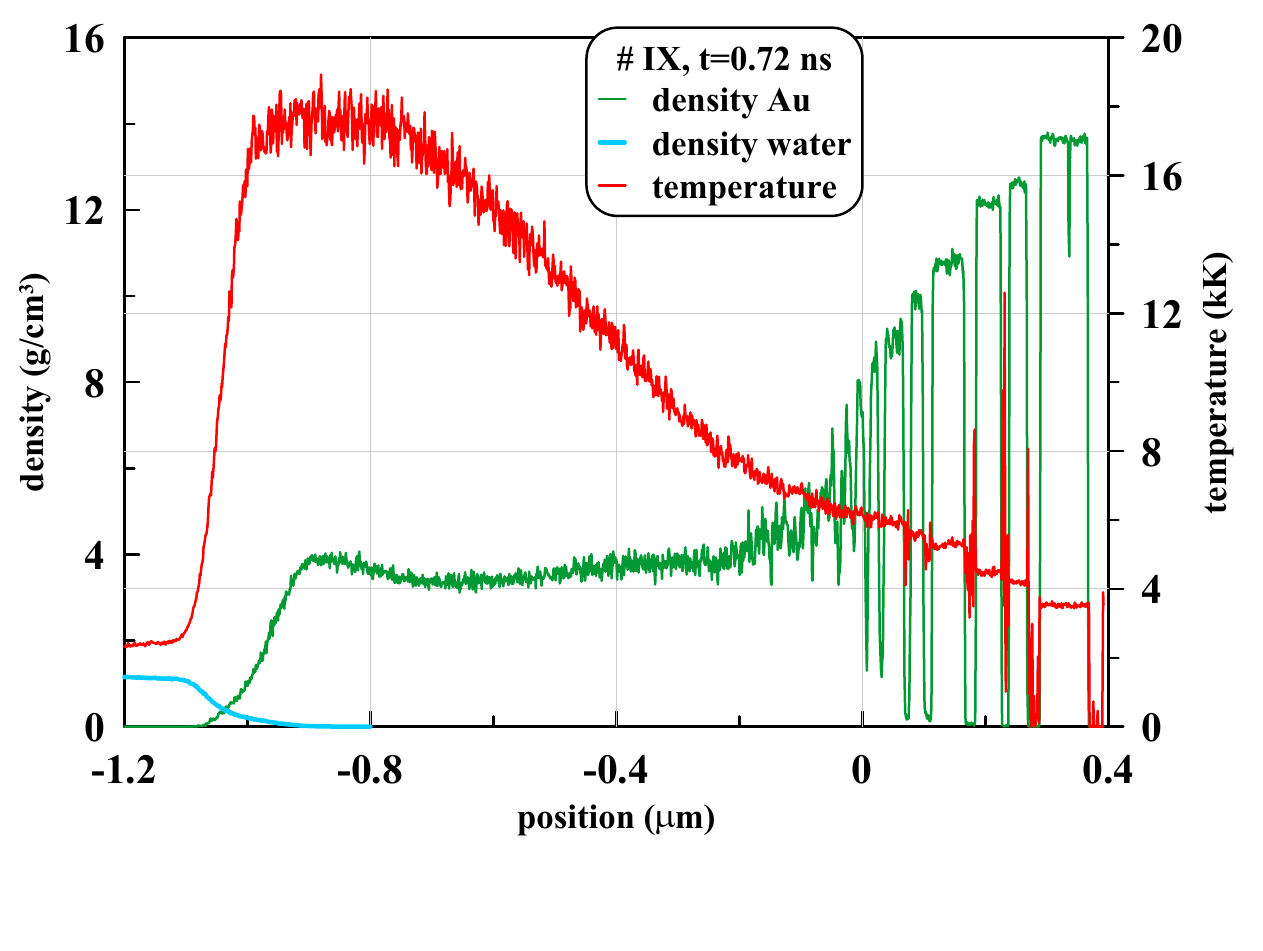}
\caption{\label{ris:11} Temperature and density profiles are shown.
Gold is strongly heated. This is a trace of a heat affected zone (HAZ) and an entropy mode.
Even foam in the tail of HAZ is hot.
It is interesting that molecular heat conduction in water cannot overcome heating of water connected with atomic diffusion of gold.
Outside the diffusion layer the water is heated thanks to thermal dissipation behind a strong shock in water;
it is strong because shock pressure is larger than bulk modulus $B\approx 1.5$ GPa of water.
     }  \end{figure}

\begin{figure}       
   \centering   \includegraphics[width=0.75\columnwidth]{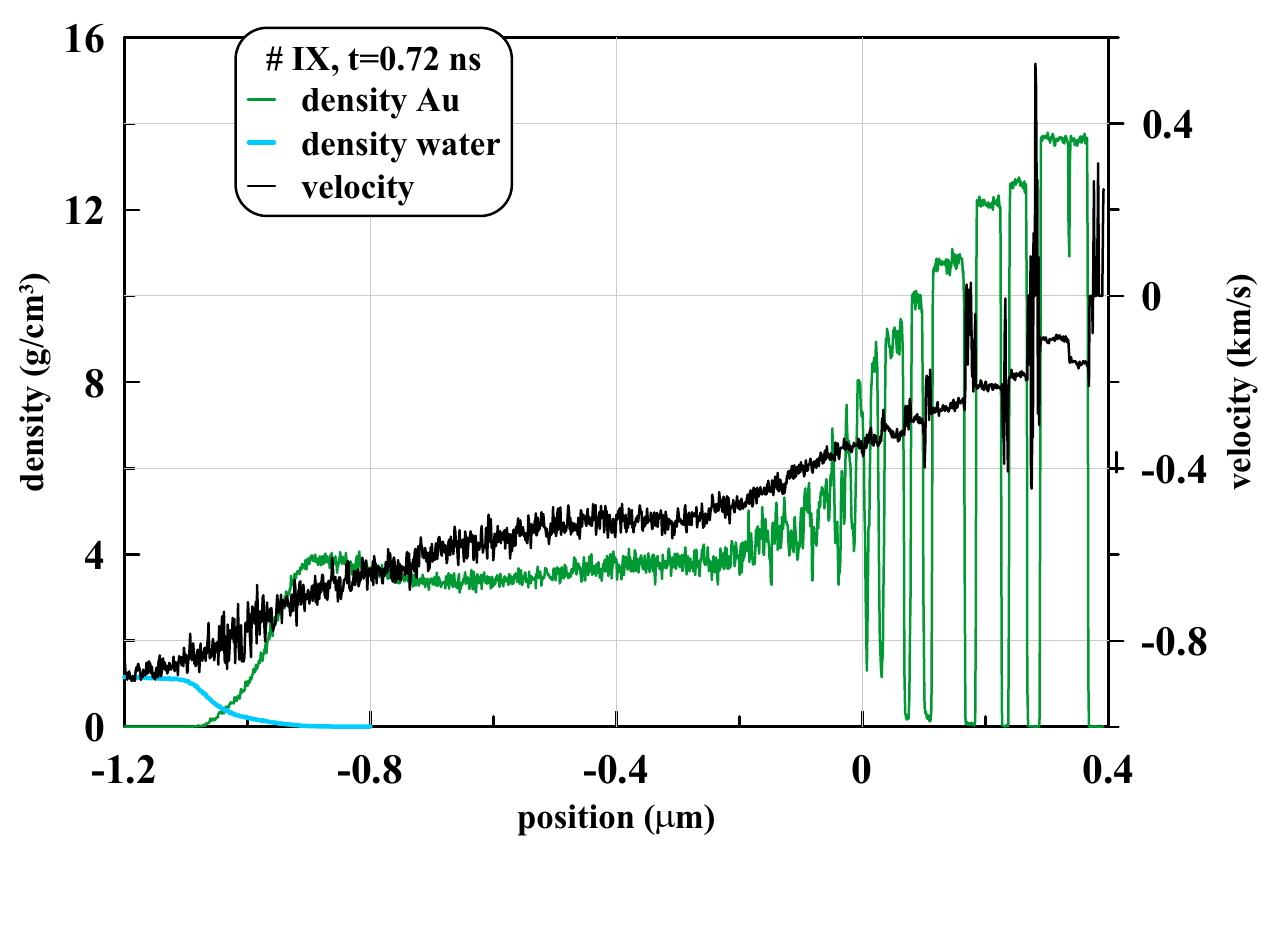}
\caption{\label{ris:12} Velocity and density profiles are shown.
Gradient of velocity is less in the region of atmosphere.
This means that matter in atmosphere moves approximately with velocity defined by the gold/water contact.
     }  \end{figure}

\begin{figure}       
   \centering   \includegraphics[width=0.75\columnwidth]{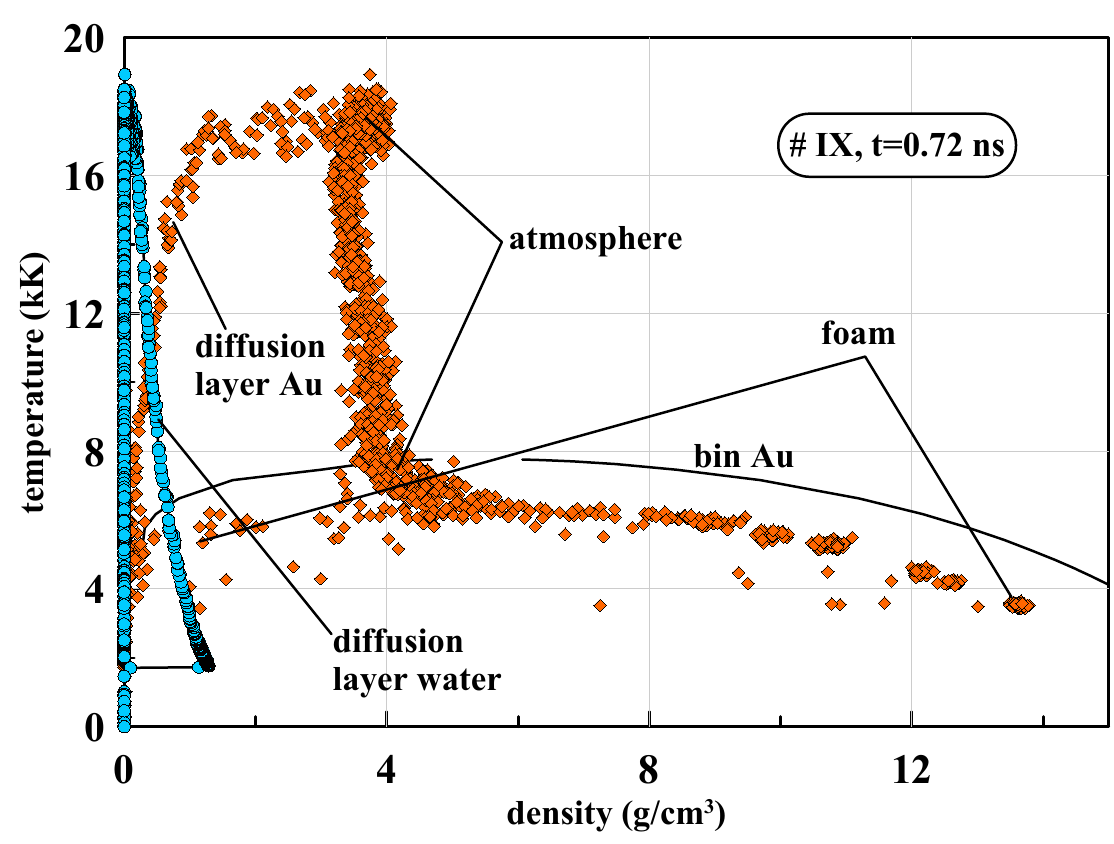}
\caption{\label{ris:13} Instant density $\rho(x,t)$ and temperature $T(x,t)$ profiles for $t=0.72$ ns plotted excluding dependence on $x$
 as a parametric function $T(\rho, t=0.72)$ at a phase diagram of gold.
   The orange rhombuses correspond to gold, the blue circles correspond to water.
     }  \end{figure}


 Fig. \ref{ris:09} shows how the instant spatial structure presented in Fig. \ref{ris:08}
   looks at the phase diagram of gold.
 Binodal (equilibrium curve), solidus, and liquidus of gold are taken according to the works
   \cite{Bushman:1993,Khishchenko2002,lomonosov_2007,rusbank1,rusbank2}.
 The binodal consists from the left and right branches relative to the critical point.
 The right branch is called a boiling curve, while the left branch is a condensation curve.


 All typical points from the structure in Fig. \ref{ris:08} are given in Fig. \ref{ris:09}.
 The AJ is in a solid state below the melting region.
 Melting is slightly higher than the melting region because the dependence $T_{melt}(p)$ from our EAM potential
  is slightly different from the curve according to equation of state from
   \cite{Bushman:1993,Khishchenko2002,lomonosov_2007,rusbank1,rusbank2}; EAM - embedded atom method.
 SW in gold and RfW follow above the mark "melting" in Fig. \ref{ris:09}.
 Pay attention to the image of the plateau "decelerated gold" from Fig. \ref{ris:08} in Fig. \ref{ris:09}.
 This is the most hot region.


 The diffusion zone follows after the plateau (we move in direction from deep gold to water).
 In this zone temperature of gold falls down to temperature of water outside the mixed zone.
 While concentration of gold atoms (the orange rhombuses in Fig. \ref{ris:09}) decreases to zero.
 The blue circles in Fig. \ref{ris:09} present water.
 The chain of them along the vertical axis with density equal to zero corresponds
  to the zero concentration of water deep into gold - see the blue curve in Fig. \ref{ris:08}
    prolonged to the right side from the mixing layer.
 The chain of the blue circles starting in the left up corner in Fig. \ref{ris:09} and going down
  increasing density and decreasing temperature relates to water in the mixing zone.
 At the left down corner the blue circles correspond to bulk water, shock in water, and shock compressed layer of water.

 \subsection{Times around nanosecond}


 The middle stage is shown in Figures \ref{ris:10}-\ref{ris:13}.
 It relates to times near one nanosecond.
 Structure of flow is described in Fig. \ref{ris:10}.
 Let's begin from deep layers of gold.
 A shock is going far away from the contact zone.
 It doesn't influence current dynamics near contact and thus isn't shown in our Figures.
 A thick layer of foam in Fig. \ref{ris:10} separates the contact from gradually solidifying continuous bulk gold.
 The melting/solidification process in continuous gold located at the bottom edge of foam
   was illustrated in Fig. \ref{ris:04}-\ref{ris:07} above.


 States of matter in foam are clear from Figures \ref{ris:11} and \ref{ris:13}.
 This is the two-phase liquid-vapor mixture occupying the both branches of the binodal curve:
   one component of the mixture is located on the boiling curve corresponding to liquid phase
    while the another one is on the condensation curve presenting saturated vapor.
 Temperature of mixture slowly gradually decreases to the right side in Fig. \ref{ris:11}.
 The bottom edge of mixture is adjoint to continuous gold while the up edge transfers gradually to the atmosphere,
   see Fig. \ref{ris:10}.


 At the left side the foam is bound by atmosphere, see Figures \ref{ris:10}-\ref{ris:13}.
 Gold in atmosphere is in overcritical states as it is shown in Fig. \ref{ris:13}.
 This is the states where thermal effects overcome still powerful cohesive properties following from interatomic interactions.
 Thus surface tension between overcritical gold and overcritical water disappears
   while their inter-diffusion is strongly enhanced.


 We call "atmosphere" the layer between the contact and foam
   because this layer is in quasi-hydrostatic equilibrium with the contact.
 This means that the layer decelerates approximately as the contact.
 Spatial gradient of velocity is decreased in the atmosphere relative to water and foam as it is shown in Fig. \ref{ris:12}.
 Deceleration of the contact at the instant $t=0.72$ ns shown is $g_{ff}\approx 1.4\cdot 10^{14}$ cm/s$\!^2.$
 When we say "atmosphere" we mean that the "free fall" deceleration $g_{ff}(t)$ changes in time more slow
   than the current acoustic time scale $t_s=h_{atm}/(c_s|_{atm}).$

\begin{figure}       
   \centering   \includegraphics[width=0.9\columnwidth]{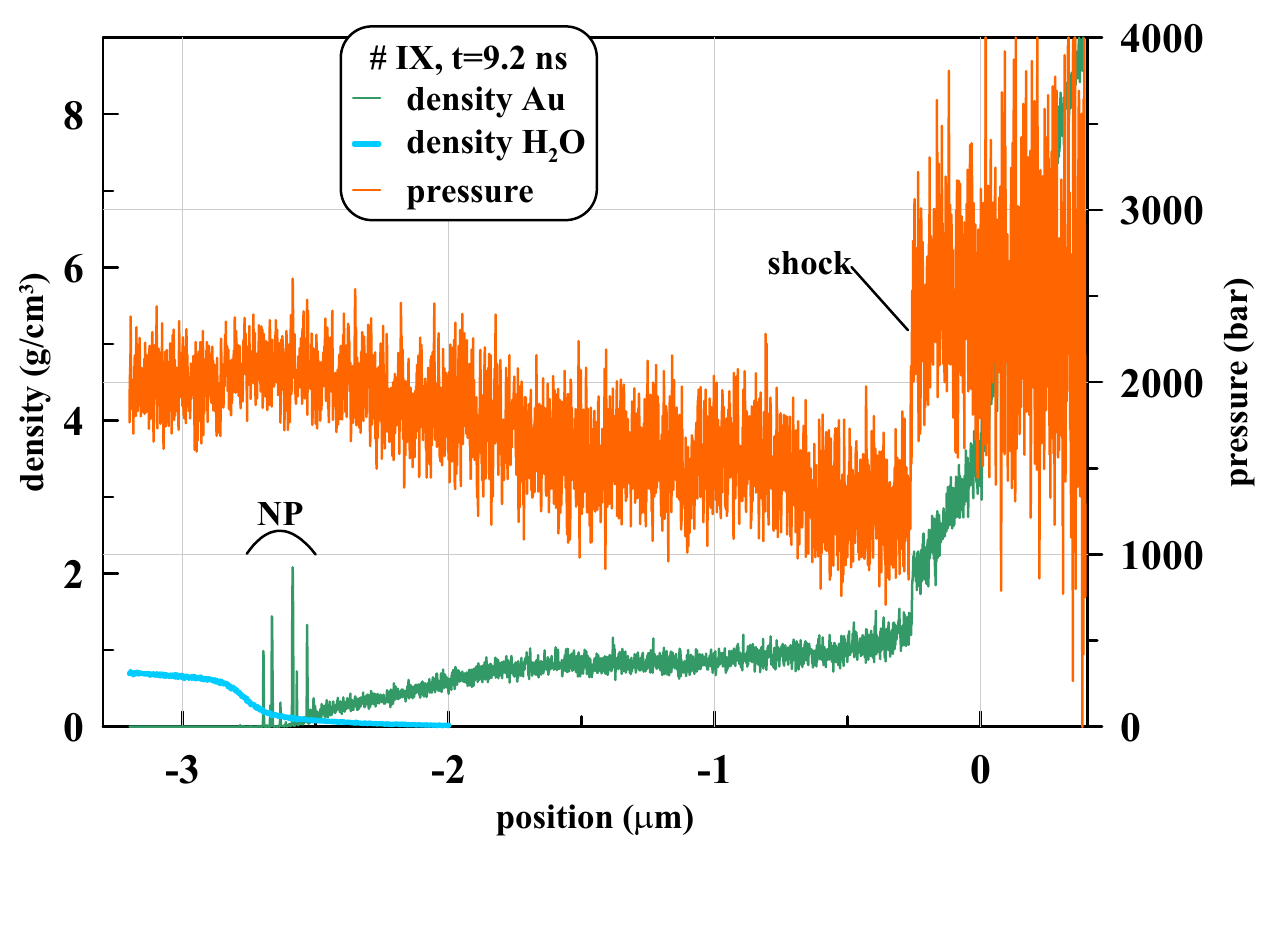}
\caption{\label{ris:14} Pressure and density profiles. The green vertical straights near the edge of the diffusive penetration of gold into water
 correspond to nanoparticles of gold. They are marked as NP.
 Weak jump at the pressure and density profiles corresponds to a shock separating back flow of gold and the rest of the motionless gold target.
 Direction of motion in the back flow is shown in Fig. \ref{ris:16} below.
     }  \end{figure}

\begin{figure}       
   \centering   \includegraphics[width=0.9\columnwidth]{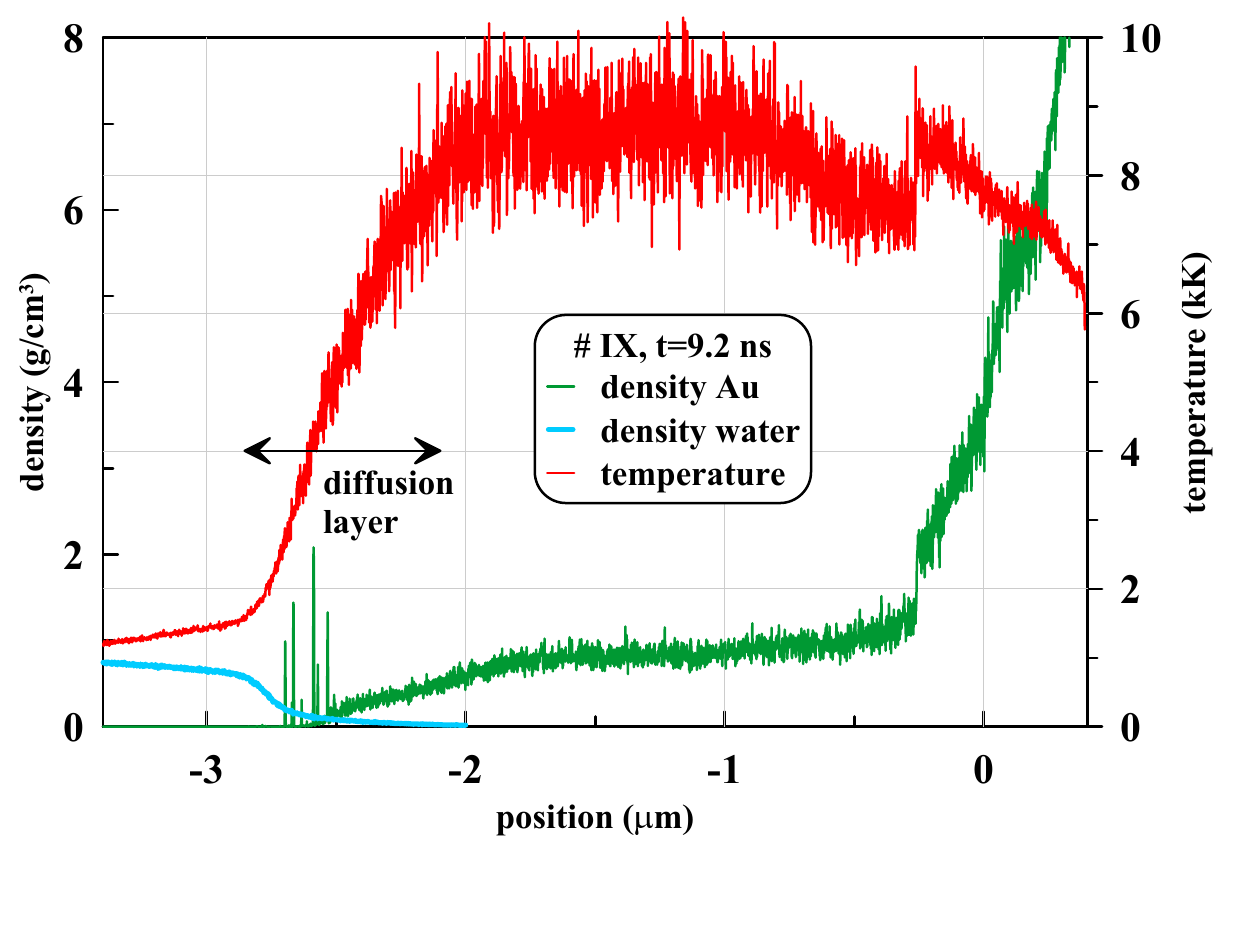}
\caption{\label{ris:15} Temperature and density profiles.
Small amount of atomic gold falls below the condensation curve and begin to condense forming nanoparticles of gold inside the diffusion layer.
Larger amount of gaseous gold with $T\approx 9$ kK and $\rho\approx 1$ g/cm$\!^3$ is above the condensation curve.
It is unclear what fraction of this gaseous gold will have time to condense before it will contact with the rest of the gold target -
direction of motion is shown in next Figure.
     }  \end{figure}

\begin{figure}       
   \centering   \includegraphics[width=0.9\columnwidth]{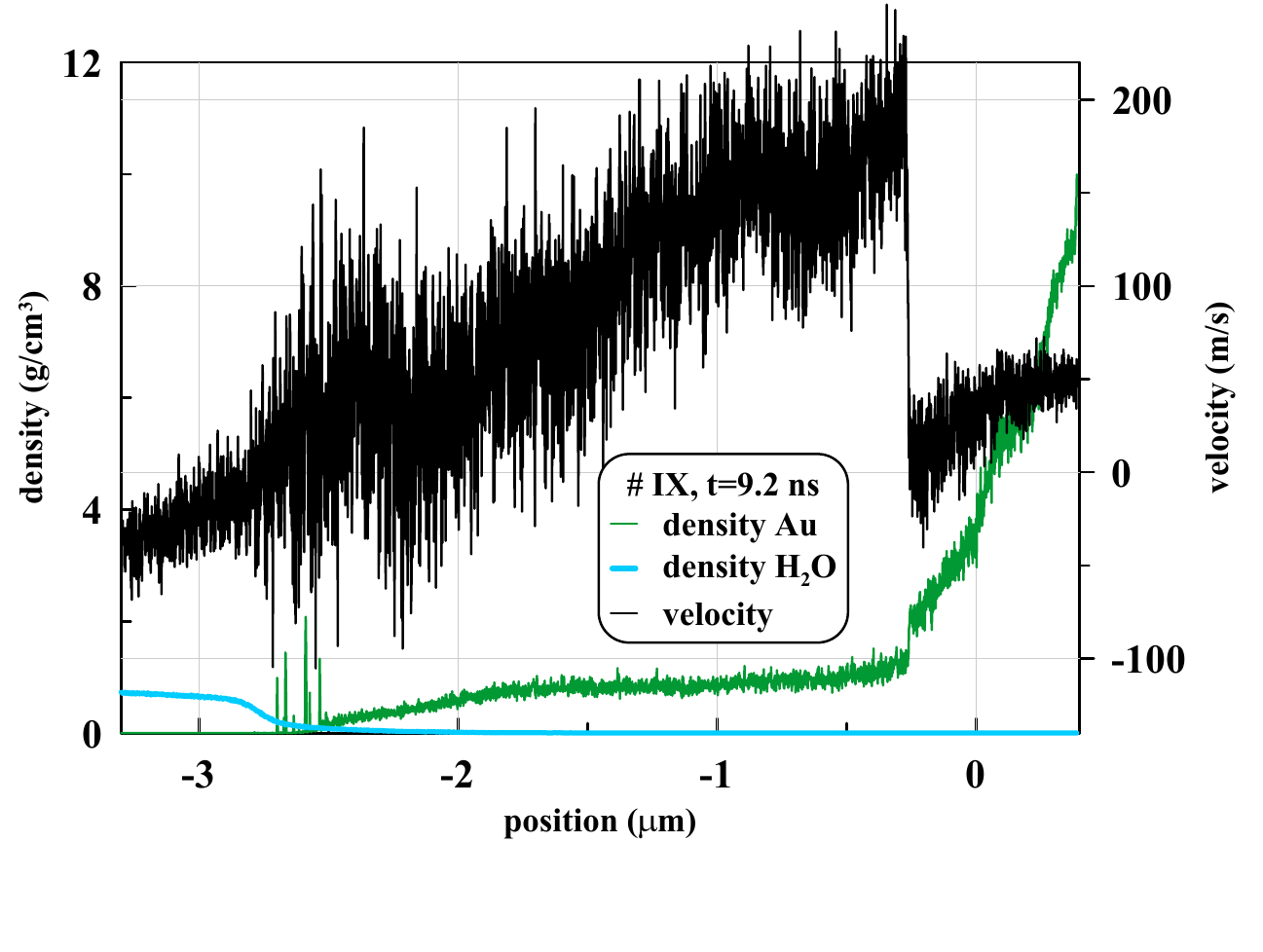}
\caption{\label{ris:16} Instant velocity profile is presented together with gold and water density profiles.
Comparison of the profiles helps to understand deep changes which take place during the time interval
 covering evolution of flow from Fig. \ref{ris:12} to this Figure.
 Motion of gold changes its direction from oriented to water to the opposite one.
 While velocity of the diffusion layer drops down to zero value - stopping of the contact zone takes place.
 Velocities of backward flow of gold achieve significant values, thus a weak shock appears in the place
   where gaseous gold meets approximately motionless, rather dense, condensed liquid.
     }  \end{figure}

\begin{figure}       
   \centering   \includegraphics[width=0.9\columnwidth]{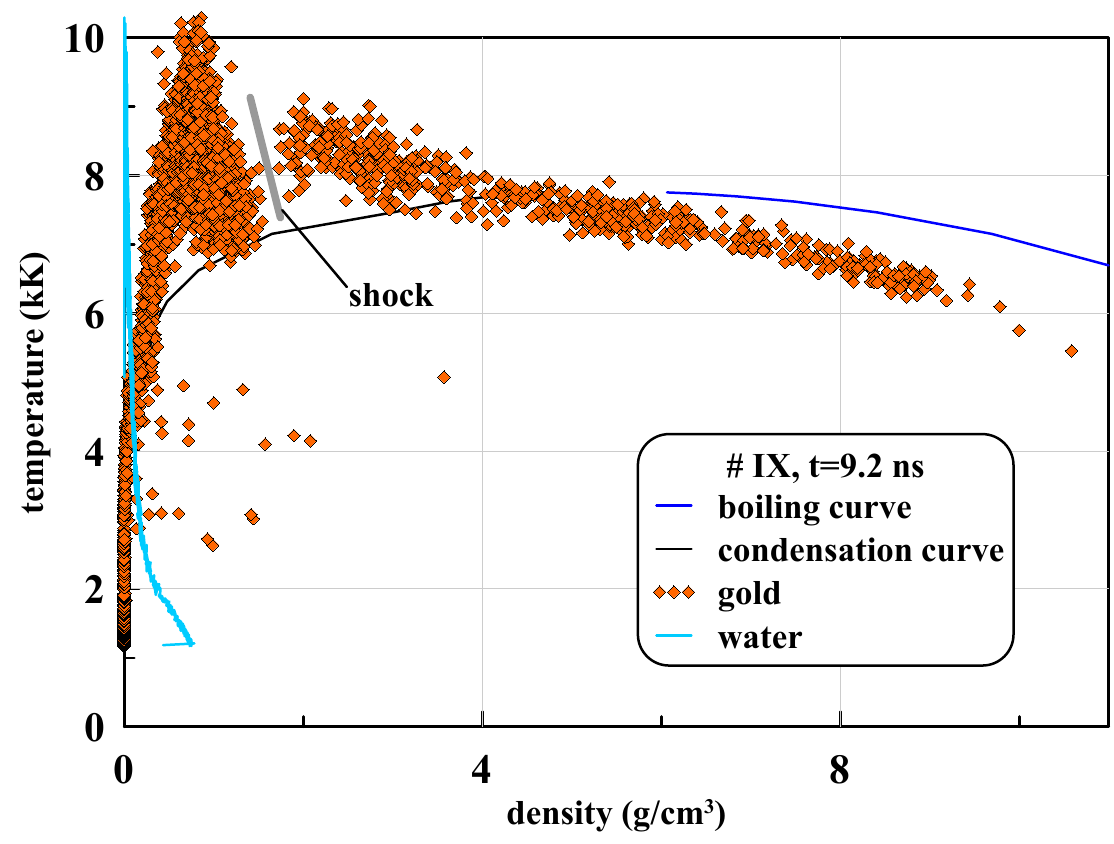}
\caption{\label{ris:17}  Instant profile $T(\rho, t=9.2 \,\rm{ns})$ at a phase diagram of gold.
   The orange rhombuses correspond to gold, the blue circles correspond to water.
   The line "shock" separates the rest of a target and the gaseous cloud moving in direction to the rest of the target, see previous Figure.
   Part of gold cooled to $T=2-4$ kK and located below the condensation curve form nanoparticles shown in Figures \ref{ris:14}-\ref{ris:16} above.
   The nanoparticles belong to the diffusion layer.
     }  \end{figure}

\begin{figure}       
   \centering   \includegraphics[width=0.9\columnwidth]{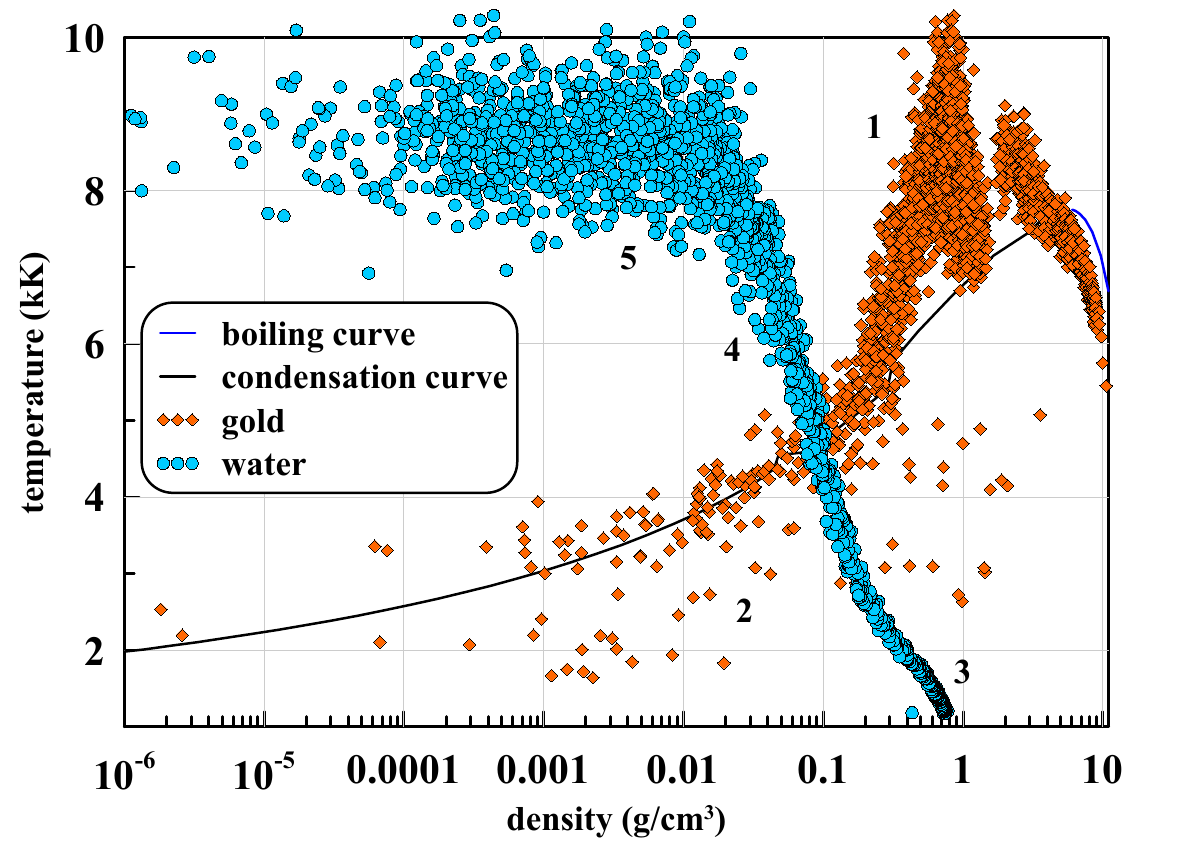}
\caption{\label{ris:18}    The same as in previous Figure but using logarithmic coordinate for density.
   This is done to see better situation at small densities.
     The orange rhombuses correspond to gold, the blue circles correspond to water.
     The digits are: 1 is gaseous gold located between the contact and the shock;
     2 is gold in the diffusion layer below the condensation curve, part of this gold condenses into nanoparticles;
     3 is water outside the diffusion zone;
     4 is water diffused into gold; 5 is negligible amount of water deep inside gold - number of circles isn't proportional to mass of water,
       mass of water presented by this group of the circles is infinitesimally small.
     }  \end{figure}


 Maximum pressure and temperature in atmosphere are 2 GPa and 18 kK, see Figures \ref{ris:10} and \ref{ris:11}.
 Thickness of atmosphere $h_{atm}$ is 0.5 micron according to Figures \ref{ris:10}-\ref{ris:12}.
 In hydrostatic equilibrium we have $h_{atm}= k_B T_{atm}/m_{Au} g_{ff},$ where $k_B$ is Boltzmann constant
   and $m_{Au}$ is mass of atom of gold.
 Taking $T=15$ kK as average temperature in the atmosphere and $g_{ff}\approx 1.4\cdot 10^{14}$ cm/s$\!^2$
   we obtain $h_{atm}=0.45$ micron.
 As deceleration $g_{ff}(t)$ decreases with time the height of the atmosphere increases
  and quasi-hydrostatic approximation loses its validity.


 Estimate of speed of sound in atmosphere based on gaseous approximation is
   $c_s|_{atm}=\sqrt{(5/3) k_B T/m_{Au}}\approx 1$ km/s for $T=15$ kK.
 Then the current acoustic time scale is $t_s=0.5$ ns.


 Atmosphere is convectively unstable because its hotter layers in an effective gravity field
   are located below the colder layers in the non-inertial frame connected with a contact.
 But it seems that there is not enough time to develop convection
   because the temporal interval of deceleration is limited to few nanoseconds
    and there are diffusion and viscosity dumping this development.
 In presence of diffusion even stronger Rayleigh-Taylor instability at the contact is suppressed.
 The Rayleigh-Taylor instability is stronger than convective instability because its increment is larger.
 The increment is larger since the density contrast at a contact is larger than the entropy contrast inside an atmosphere.


 We don't see indications of development of convection in our molecular dynamics simulation \# VIII listed in Table 1.
 Lateral size is large in this simulation.
 The simulation \# VIII is specially intended to follow development of the multi-dimensional instability.
 Even the Rayleigh-Taylor instability very weakly develops in the situation corresponding to simulations \#\# VIII and IX.


 Velocity of the contact zone is $\approx 800$ m/s at the instant $t=0.72$ ns presented in Figures \ref{ris:10}-\ref{ris:13}, see Fig. \ref{ris:12}.
 The contact is smeared thanks to diffusion.
 Densities of gold and water are shown separately in Figures \ref{ris:10}-\ref{ris:12}.
 Expansion velocity of the diffusion zone (from one edge to another one, concentrations change from 0\% to 100\%) thanks to difference of hydrodynamic velocities at the edges
  is 100 m/s at this instant.
 Expansion velocity of a gold/water mixture relative to matter due to diffusion is $(d/dt)2\sqrt{D \, t}=12$ m/s for $t=0.72$ ns and diffusion coefficient $D=0.001$ cm$\!^2/$s.

 \subsection{Times around 10 nanoseconds}


 Qualitative changes in structure of flow take place during the time interval between $t\sim 1$ ns and $t\sim 10$ ns.
 Structure described above in Figures \ref{ris:10}-\ref{ris:13} corresponding to the stage $t\sim 1$ ns
     was supported thanks to momentum and kinetic energy of atmosphere and foam.
 This momentum was directed to the water side, see Fig. \ref{ris:12}.
 Water resists to expansion of gold, thus gradually through this resistance the momentum accumulated in the gold atmosphere and foam
    is transferred to the shock compressed layer of water.
 Amount of momentum is finite.
 Therefore it is not surprising that there is a stage when all momentum of gold directed to water is exhausted.


 After that pressure at the contact zone is supported by gas and saturation pressure of hot enough gold.
 Momentum transfer and resistance of water maintain the "free fall" or quasi-gravity like
   (thanks to the Einstein principle of equivalence of gravitational and inertial mass) deceleration $g_{ff}$ of a gold/water contact.
 In turn the deceleration $g_{ff}$ leads to creation of the atmospheric like quasi-hydrostatic layer of gold decelerated by the contact.


 Deceleration leads to stopping of the contact.
 Velocity of the contact decreases to zero value and after that changes sign - slow back motion of gold begins.
 At the same stage the deceleration $g_{ff}$ drops down to small values.
 Thus thickness of the corresponding "atmosphere" and acoustic time scale for sound to pass atmosphere become large,
   larger than spatial and temporal scales related to current motion.
 Then the atmosphere as a significant element of the structure disappears.


 Returning to the initial stages we have to mention that the opposite directed momentums in gold appear after laser heating
   and creation of the heat affected zone (HAZ) in gold.
 The positive momentum (positive velocity is directed along the $x$-axis, $x$ is growing into bulk gold)
   is taken away by the shock running into bulk gold.
 The negative momentum is initially accumulated in the near contact layer of gold expanding to water.
 As was said, due to resistance of water the negative momentum transits to the shock in water and to the shock compressed layer of water.


 Figures \ref{ris:14}-\ref{ris:18} present the situation corresponding to the transient stage.
 This is the transient from the momentum transfer stage to the stage when pressure at the contact zone
   is supported by gas and vapor pressure of slowly cooling gold.
 Gold slowly cools because thermal conductivity in gaseous and two-phase (liquid-vapor) gold is much less than in condensed phase.


 Pressure of gold near contact drops down approximately ten times during the time interval from 0.72 ns to 9.2 ns,
   compare Figures \ref{ris:10} and \ref{ris:14}.
 Gradient of pressure in gold layer near contact decreases $\approx 10^2$ times from 2 GPa per 0.7 $\mu$m to 800 bar per 2.5 $\mu$m.


 Density decreases from 4 g/cm$\!^3$ to approximately 1 g/cm$\!^3,$ compare Figures \ref{ris:10}-\ref{ris:13} and \ref{ris:14}-\ref{ris:18}.
 Geometrical thickness of the near contact gold layer increases approximately 3.5 times.


 Decrease of deceleration $g_{ff}$ (while temperatures decreases much more slowly and remains high at the stage $t\sim 10$ ns)
   causes increase of height of atmosphere $h_{atm}$.
 This process looks like a back flow or outflow of gaseous gold from a contact.
 Velocity of gold turns back from the direction oriented to water side.
 To quantify the scale of this reorientation let's compare Figures \ref{ris:12} and \ref{ris:16}.

\begin{figure}       
   \centering   \includegraphics[width=0.7\columnwidth]{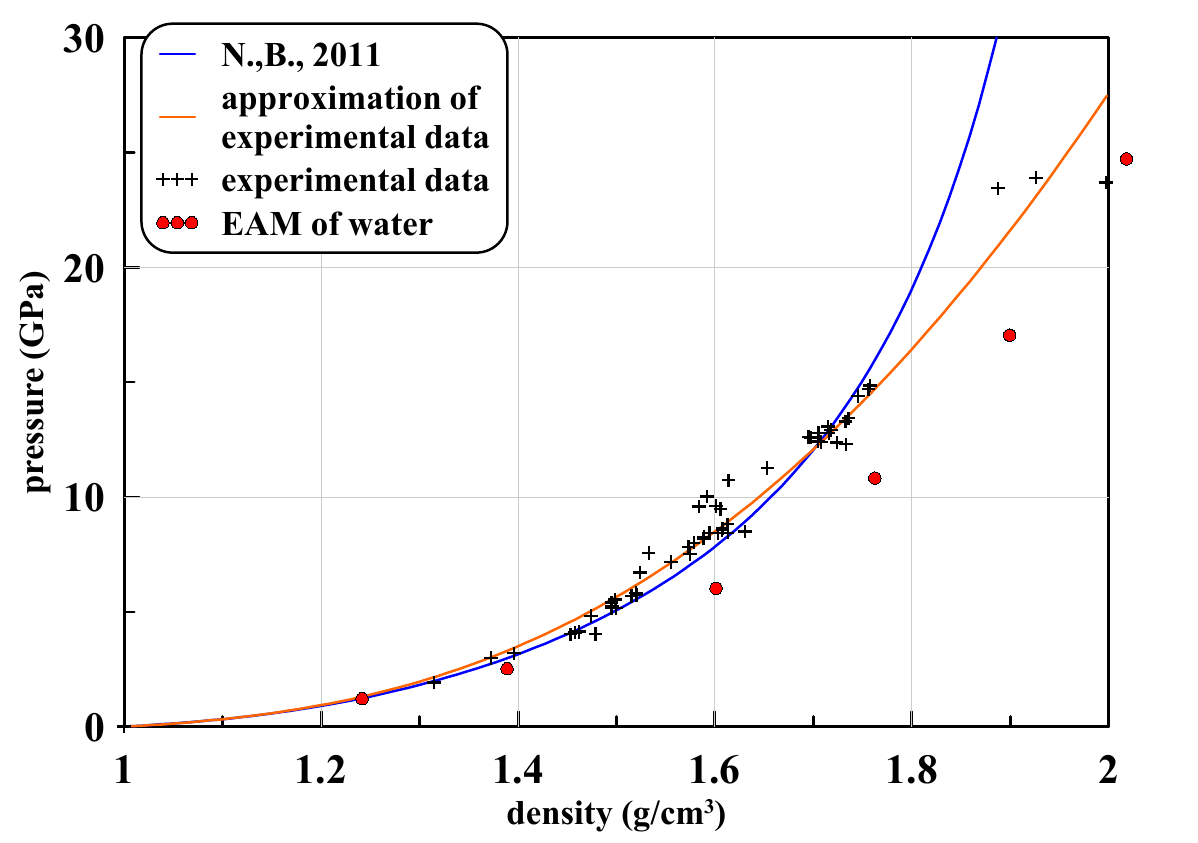}
\caption{\label{ris:19}    Comparison of EoS \cite{NB:2011} with experimental data on shock compression from \cite{rusbank1,rusbank2}.
We also compare with experiment the Hugoniot adiabatic curve following from the EAM potential for water used in simulations \#\# VIII, IX.
The abbreviation N.,B., 2011 relates to the paper \cite{NB:2011}.
Semi-analytic approximation of experimental data is taken from \cite{INA.jetp:2018.LAL,INA.arxiv:2018.LAL}.
It is valid from infinitesimal compressions and up to 120 GPa - the highest measured pressures.
     }  \end{figure}

\begin{figure}       
   \centering   \includegraphics[width=0.7\columnwidth]{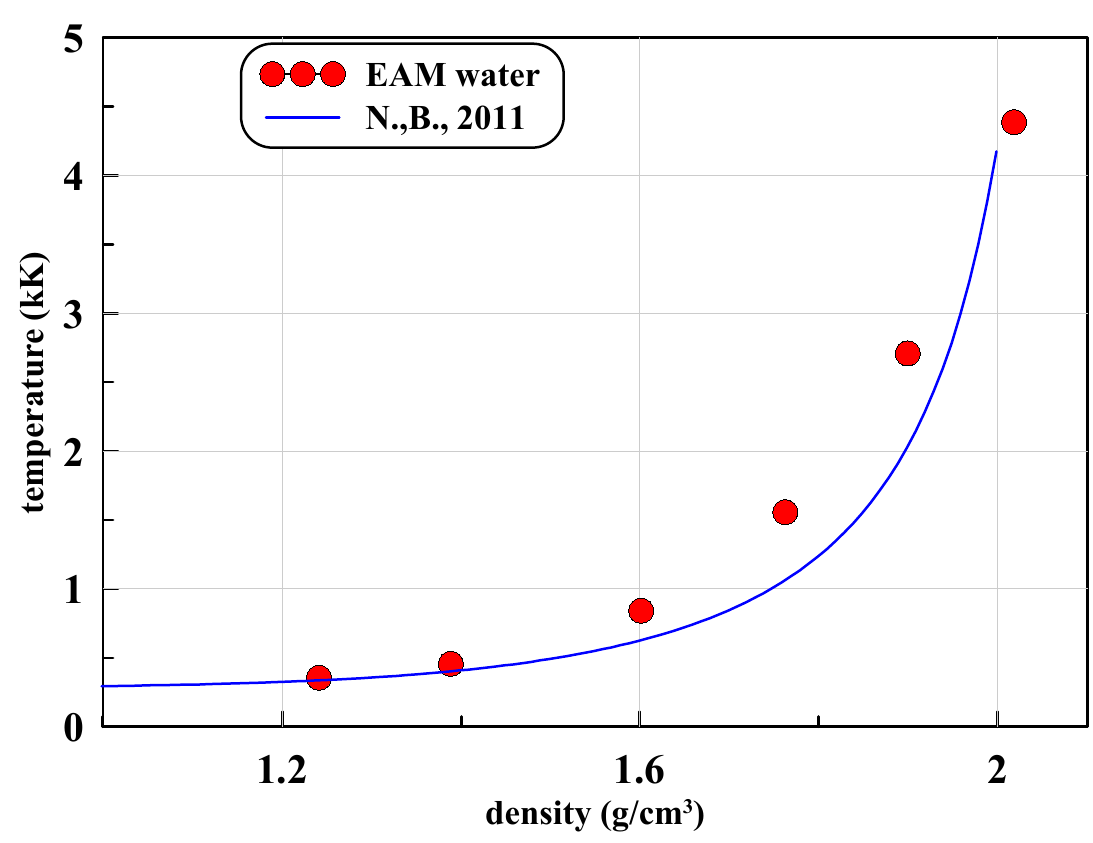}
\caption{\label{ris:20}   Heating of water by shock is shown.
     }  \end{figure}


 The outflow of gold achieves significant velocities exceeding local speed of sound.
 Thus a weak shock forms in the place where the outflow impacts dense molten motionless gold
   belonging to the right branch (boiling curve) of a binodal (coexistence or phase equilibriun curve), see Fig. \ref{ris:17};
     position of the shock is marked in Fig. \ref{ris:14}.
 According to Fig. \ref{ris:16} the velocity jump at the weak shock seen in Figures \ref{ris:14}-\ref{ris:16} is near 200 m/s.
 Gas like estimate of speed of sound $c_s=\sqrt{(5/3) k_B T/m_{Au}}$ gives 730 m/s for $T=7.5$ kK;
   this is temperature in gaseous gold near the left side of the jump, see Fig. \ref{ris:15}.
 The estimate based on the slope of the condensation curve $c_s=\sqrt{\Delta p/\Delta\rho}$ gives 390 m/s;
   here the differences $\Delta p$ and $\Delta\rho$ are taken along the condensation curve.
 Speed of sound from EAM (embedded atom method) interatomic potential of gold used in molecular dynamics simulations \#\# VIII, IX
   should be less than 200 m/s in the conditions corresponding to the state near the jump.


 Expansion of gold, decrease of its density and temperature mainly due to diffusive mixing with much colder water
      lead to condensation of atomic gold in clusters and nanoparticles.
 Thanks to cooling the gaseous gold intersects the condensation curve shown in Figures \ref{ris:17} and \ref{ris:18}.
 Nanoparticles are seen at the density profile of gold in Figures \ref{ris:14}-\ref{ris:16}.
 They are marked by "NP" in Fig. \ref{ris:14}.
 The nanoparticles begin to appear after $t\sim 1$ ns.
 Their amount and size grow with time.


 Open question remains about total number of nanoparticles produced by a laser pulse.
 Indeed, there are significant quantity of gaseous gold above the condensation curve in Figures \ref{ris:17} and \ref{ris:18}.
 They correspond to the cloud located between the contact and the jump in Figures \ref{ris:14}-\ref{ris:16}.
 But it is unclear, how much of them will have time to condense before they collide with continuous gold corresponding to the rest of a gold target.
 Even longer simulations are planned to address this problem.
 Nevertheless, it is obvious that the nanoparticles mixed with water (they are seen in Figures \ref{ris:14}-\ref{ris:16})
   have not a chance to return to the gold target.





\begin{figure}       
   \centering   \includegraphics[width=0.7\columnwidth]{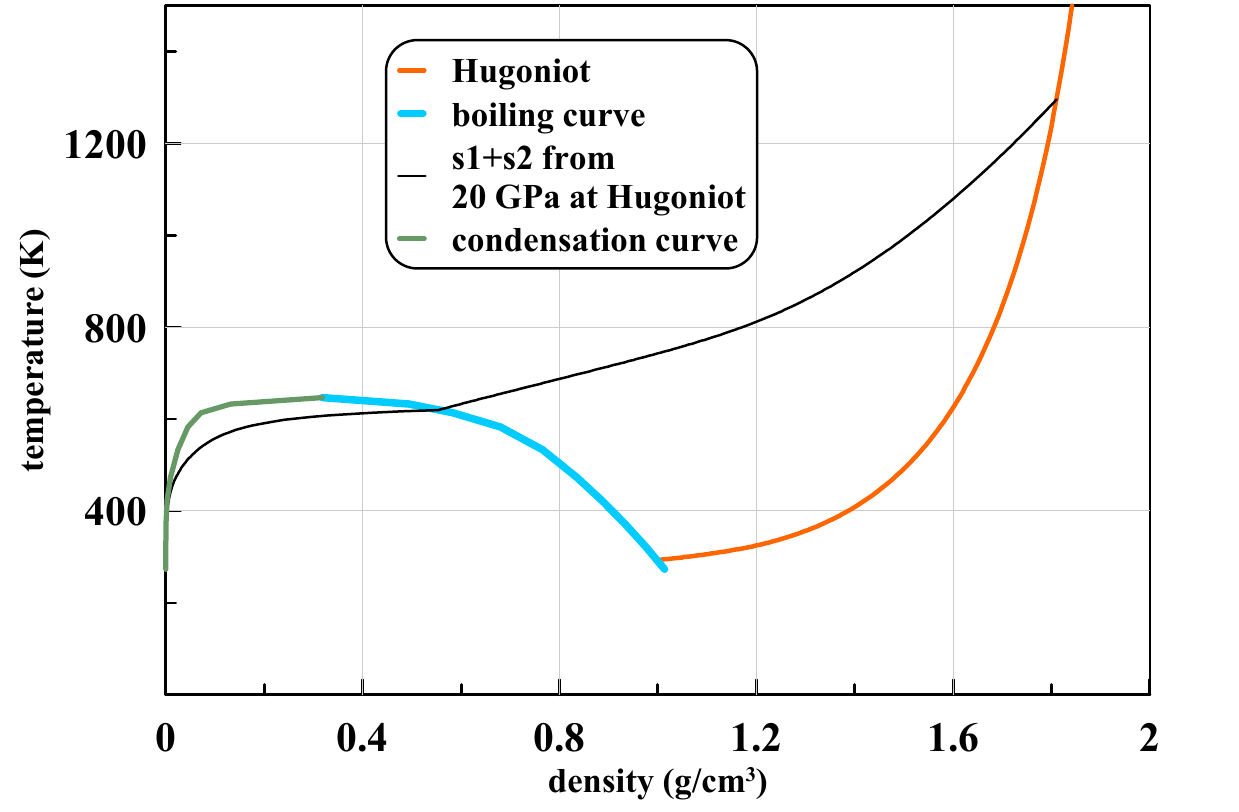}
\caption{\label{ris:21}   Expansion of water from a shock compressed state down to binodal where behavior of water changes
  from the condensed phase behavior to gaseous like behavior.
  Expansion is accompanied with decrease of temperature spent for mechanical work against pressure.
  The adiabatic curve consists from two parts s1 and s2.
  The first part s1 corresponds to the one-phase interval.
  It covers the way from the initial point at the Hugoniot adiabatic curve to the intersection point between the adiabatic curve and binodal.
  The second part s2 covers expansion inside the two-phase region.
  Calculations are made using EoS \cite{NB:2011}.
  In the initial point pressure is 20 GPa.
     }  \end{figure}

\begin{figure}       
   \centering   \includegraphics[width=0.7\columnwidth]{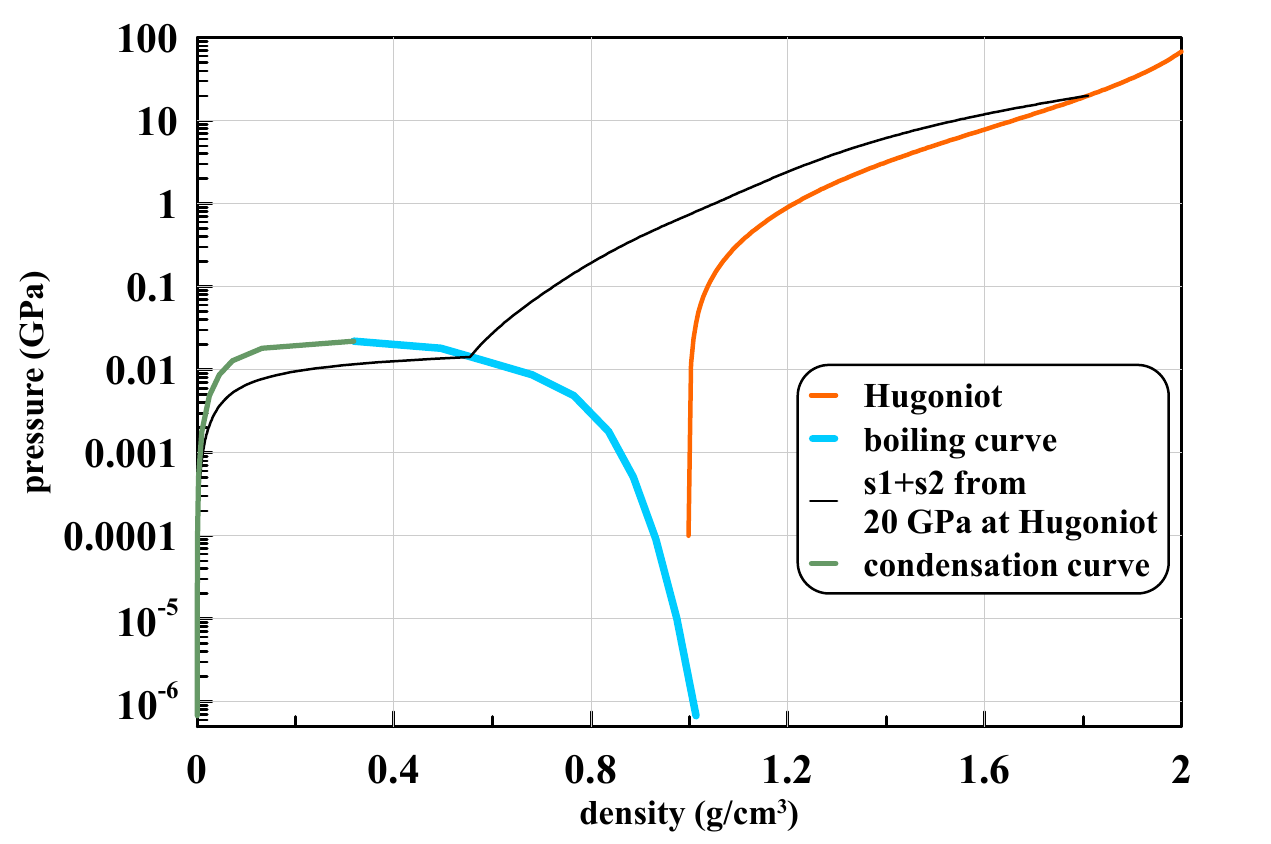}
\caption{\label{ris:22}   Here the adiabatic curve s1+s2 from previous Figure is shown at the density-pressure phase diagram.
During expansion to atmospheric pressure a water layer passes through many orders of magnitudes at the pressure axis.
     }  \end{figure}

\begin{figure}       
   \centering   \includegraphics[width=0.7\columnwidth]{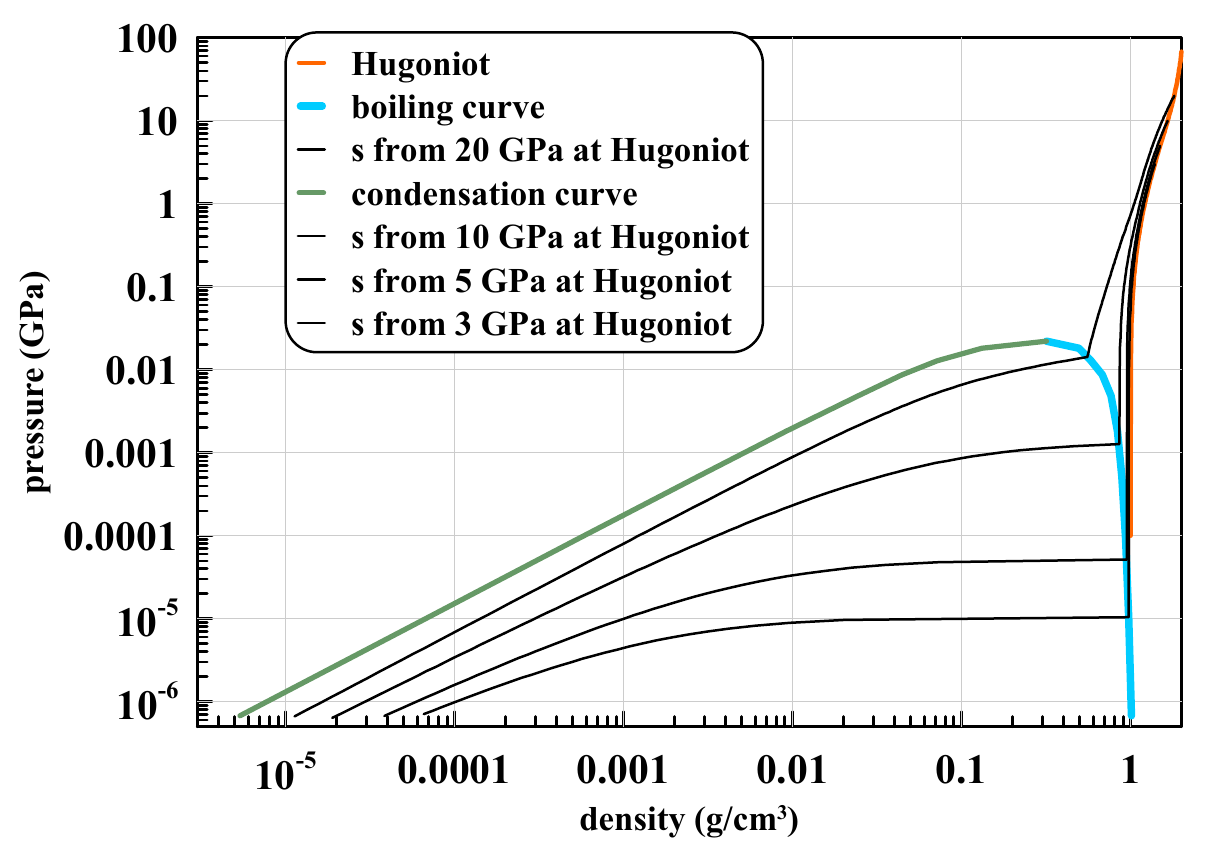}
\caption{\label{ris:23}   This is the plot taken from previous Figure but now in double logarithmic scales.
This is necessary to see how large is expansion degree at pressures equal to one bar and below.
     }  \end{figure}

\begin{figure}       
   \centering   \includegraphics[width=0.7\columnwidth]{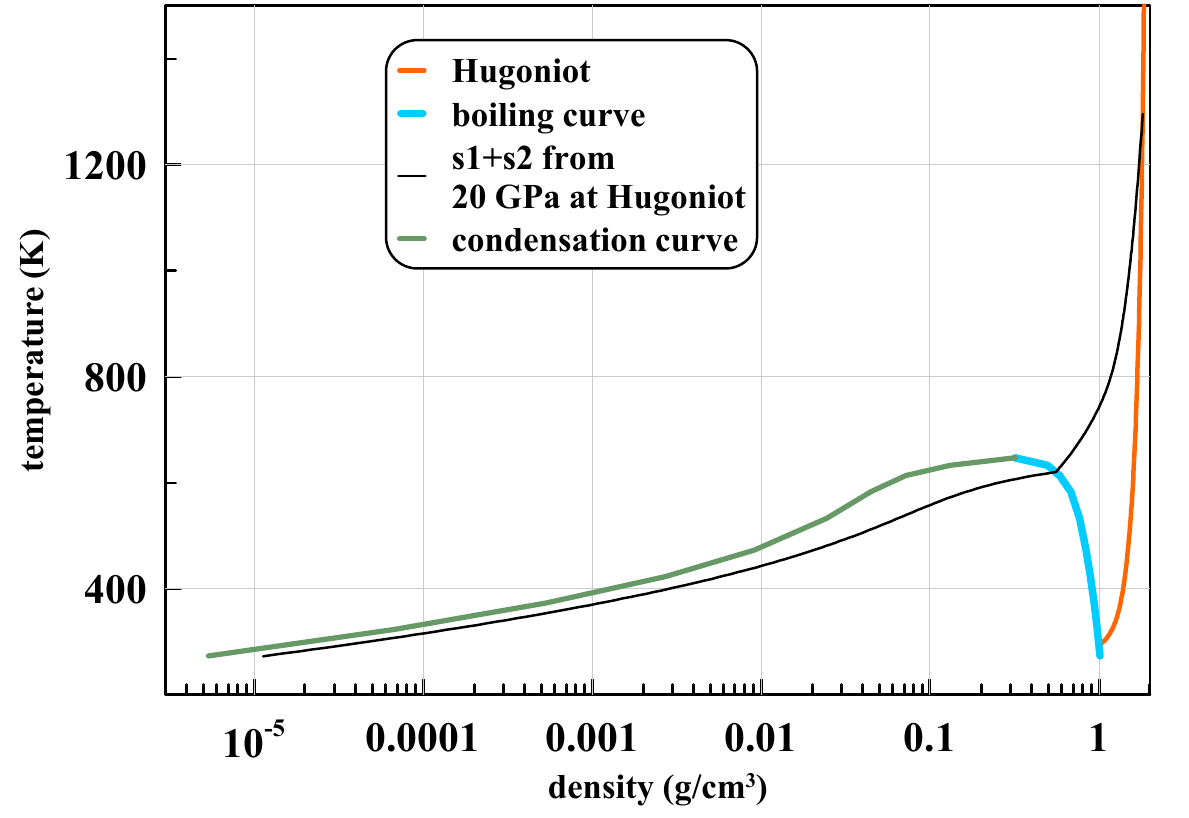}
\caption{\label{ris:24}   Temperature decreasing as expansion proceeds is illustrated.
 Water is shock heated to $\approx 1.3$ kK.
 During expansion 6000 times to pressure 0.1 bar temperature drops down to $\approx 320$ K.
     }  \end{figure}


   \section{Heating of water and bubble formation}

 \subsection{Equation of state for water and shock adiabat}


 We have tried three equation of states (EoS) for water. They are: van der Waals EoS, Tait EoS,
   and semi-analytical EoS developed by Nigmatulin and Bolotnova \cite{NB:2011}.
 It is necessary to describe (a) Hugoniot adiabatic curve,
 (b) binodal of water,
 (c) one-phase adiabatic curves passing from the Hugoniot adiabat to binodal,
  and (d) two-phase adiabats presenting prolongations of the one-phase adiabats
     through intersection with binodal into two-phase vapor-liquid mixture.
 Thus rather wide range of a phase diagram should be covered; excluding solid states.
 The standard van der Waals EoS is relevant around a two-phase region.
 But it has rather large excluded volume and cannot be applied for approach to the Hugoniot curve
   with significant compressions of liquid water from normal conditions.
 The Tait EoS is better near Hugoniot but it difficult to connect Hugoniot and binodal using it.
 The EoS \cite{NB:2011} is valid in the region of a phase diagram from Hugoniot to binodal.


 Shock adiabat of water starting from normal state according to measurements \cite{rusbank1,rusbank2} is shown in Fig. \ref{ris:19}
   together with shock adiabats used in our calculations.
 We see that EoS \cite{NB:2011} may be used for compressions below 17 GPa        
  and has right speed of sound in an acoustic limit.
 The EAM (embedded atom method) interatomic potential of water developed by Zhakhovsky
   gives right approximation of the Hugoniot curve in a wide range of pressures up to the Megabar region;
     this potential is presented in https://www.researchgate.net/project/Development-of-interatomic-EAM-potentials.


 The range of pressures up to 17-18 GPa, where EoS \cite{NB:2011} is valid, is enough for our application
   with gold ablation into water.
 Indeed, acoustic impedance of water is low relative to the impedance of gold.
 Therefore even the most powerful laser impact with ultrashort duration corresponding to simulations \#\# VIII and IX
   from Table 1
     produces moderate compressions in water, see Fig. \ref{ris:08}.
 Pressures in gold are above Megabar while in water they are below 20 GPa.
 For laser actions with smaller absorbed energy $F_{abs}$ or longer durations the pressures in water are smaller.


 To address the important problem of formation of a bubble filled with water vapor
   we have to know the thermal history of the water layers adjoining to the contact with gold.
 There are two sources of heating of water.
 First it is heated by dissipation of kinetic energy thanks to friction behind a front of a shock wave.
 The second source is linked to conductive heating of water from hot gold.
 In Fig. \ref{ris:20} the Hugoniot curves are shown at the density-temperature plane.
 We see that the shock heating is satisfactory described.
 Comparison with experimental temperature measurements is also given in paper \cite{NB:2011}.
 It shows that up to a few kK the EoS \cite{NB:2011} correctly presents shock heating.

 \subsection{Transition from strongly overcritical states to huge rarefactions of water}


 Let's consider first the dissipation in shock ignoring thermal conduction.
 Then during expansion water cools down along an adiabatic curve; no thermal exchange with surrounding medium.
 Example of this curve is shown in Fig. \ref{ris:21}.
 The adiabatic curve is separated to two parts: one (s1) along condensed phase, while another (s2) in vapor-liquid mixture.


 The adiabatic curve shown in Fig. \ref{ris:21} starts from the Hugoniot curve for water.
 Pressure behind shock front in this point is 20 GPa.
 This is the highest possible pressure in water in the case of gold ablation in water.
 Nevertheless the adiabatic curve in Fig. \ref{ris:21} crosses the binodal at its right branch corresponding to boiling curve,
   thus below the critical point; entropy is less than entropy at the adiabat passing the binodal at the critical point.


 This limits expansion of water when pressure drops down to 0.1-1 bar level, see Fig. \ref{ris:22}.
 The adiabatic curves starting from Hugoniot adiabatic curve at 20, 10, 5, and 3 GPa are shown in Fig. \ref{ris:23}.
 Expansion of water along the strongest adiabatic curve beginning from 20 GPa and finishing at 0.1 bar
   gives increase in volume 6000 times
     relative to volume of water in normal conditions before the laser action.
 While for adiabatic curve starting from the Hugoniot curve at 3 GPa the expansion degree at $p=0.1$ bar
    is only $\approx 10,$ see Fig. \ref{ris:23}.
 This means that the water bubble is filled with water vapor corresponding to different entropy values.
 The coldest layers form the outer shells of a bubble.


 Cooling history and temperature decrease is shown in Fig. \ref{ris:24}.
 Temperature decreases approximately twice during three times expansion along the one-phase interval s1 of the adiabatic curve.
 After that at the two-phase interval s2 the temperature again decreases twice but now expansion is large: $\approx 2500$ times.

   \section{Conclusion}

 Ablation of gold in water is considered in different regimes of laser action.
 We vary absorbed energy and duration of a pulse.
 Structure of ablation flow and its structural reconstruction during evolution are analyzed.
 It is shown that strong pulse transfers gold into overcritical states.
 Water with its low critical parameters for long time exists in its overcritical states.
 Hot compressed water and gold actively interpenetrate each other through diffusion process.
 Rather thick mixed contact layer is formed.

 Temperature decreases in this layer from large temperature in gold to relatively small values in water.
 Gold is rather cold near the water edge of the mixed zone.
 Thus gold crosses condensation curve and formation of clusters and nanoparticles begins.
 These nanoparticles are mixed with water, thus they cannot stick back to remnants of a bulk target
   and remain in water during any extent of water expansion.

 Expansion, pressure and temperature drops during water expansion are considered.

 Acknowledgement. This work was supported by the Russian Science Foundation grant 14-19-01599.

\bibliographystyle{elsarticle-num}

\end{document}